\pdfpageattr {/Group << /S /Transparency /I true /CS /DeviceRGB>>}
\documentclass[12pt,aps,reprint,prl,superscriptaddress,longbibliography]{revtex4-1}
\usepackage[version=3]{mhchem}
\usepackage{graphicx}
\usepackage{floatrow}
\usepackage{amssymb}
\usepackage{amsmath}
\usepackage{commath}
\usepackage{blindtext}
\usepackage{siunitx}
\usepackage{miller}
\usepackage{xcolor}
\usepackage{float}
\usepackage{lipsum}
\usepackage{bm}
\usepackage{ulem}
\usepackage{physics}
\usepackage{pdfpages}
\setcitestyle{super}

\usepackage{etoolbox}

\makeatletter
\patchcmd{\frontmatter@RRAP@format}{(}{}{}{}
\patchcmd{\frontmatter@RRAP@format}{)}{}{}{}
\renewcommand\Dated@name{Dated: }
\patchcmd{\@outputpage@head}{\@ifx{\LS@rot\@undefined}{}{\LS@rot}}{}{}{}
\makeatother

\begin{document}

\title{Berry phase engineering at oxide interfaces}

\author{Dirk J.\ Groenendijk}
\email{d.j.groenendijk@tudelft.nl}
\affiliation{Kavli Institute of Nanoscience, Delft University of Technology, P.O. Box 5046, 2600 GA Delft, Netherlands}

\author{Carmine Autieri}
\email{equal contribution}
\affiliation{Consiglio Nazionale delle Ricerche CNR-SPIN, Italy}
\affiliation{International Research Centre MagTop, Institute of Physics, Polish Academy of Sciences, Aleja Lotnik\'ow 32/46, PL-02668 Warsaw, Poland}

\author{Thierry C.\ van Thiel}
\email{equal contribution}
\affiliation{Kavli Institute of Nanoscience, Delft University of Technology, P.O. Box 5046, 2600 GA Delft, Netherlands}

\author{Wojciech Brzezicki}
\affiliation{Consiglio Nazionale delle Ricerche CNR-SPIN, Italy}
\affiliation{International Research Centre MagTop, Institute of Physics, Polish Academy of Sciences, Aleja Lotnik\'ow 32/46, PL-02668 Warsaw, Poland}

\author{Nicolas Gauquelin}
\affiliation{Electron Microscopy for Materials Science (EMAT), University of Antwerp, 2020 Antwerp, Belgium}

\author{Paolo Barone}
\affiliation{Consiglio Nazionale delle Ricerche CNR-SPIN, Italy}

\author{Karel H.\ W.\ van den Bos}
\author{Sandra van Aert}
\author{Johan Verbeeck}
\affiliation{Electron Microscopy for Materials Science (EMAT), University of Antwerp, 2020 Antwerp, Belgium}

\author{Alessio Filippetti}
\affiliation{Dipartimento di Fisica, Universit\`a  di Cagliari, Cagliari, Monserrato 09042-I, Italy}
\affiliation{CNR-IOM, Istituto Officina dei Materiali, Cittadella Universitaria, Cagliari, Monserrato 09042-I, Italy}

\author{Silvia Picozzi}
\affiliation{Consiglio Nazionale delle Ricerche CNR-SPIN, Italy}

\author{Mario Cuoco}
\affiliation{Consiglio Nazionale delle Ricerche CNR-SPIN, Italy}
\affiliation{Dipartimento di Fisica ``E.\ R.\ Caianiello'' Universit\`a degli Studi di Salerno, 84084 Fisciano, Italy}

\author{Andrea D.\ Caviglia}
\email{a.caviglia@tudelft.nl}
\affiliation{Kavli Institute of Nanoscience, Delft University of Technology, P.O. Box 5046, 2600 GA Delft, Netherlands}

\maketitle


\textbf{
Geometric phases in condensed matter play a central role in topological transport phenomena such as the quantum, spin and anomalous Hall effect (AHE)~\cite{Berry45, von1986quantized, jungwirth2012spin, karplus1954hall, nagaosa2010anomalous}. In contrast to the quantum Hall effect---which is characterized by a topological invariant and robust against perturbations---the AHE depends on the Berry curvature of occupied bands at the Fermi level and is therefore highly sensitive to subtle changes in the band structure~\cite{xiao2010berry, nagaosa2010anomalous}. A unique platform for its manipulation is provided by transition metal oxide heterostructures, where engineering of emergent electrodynamics becomes possible at atomically sharp interfaces~\cite{zubko2011interface, hwang2012emergent}. We demonstrate that the Berry curvature and its corresponding vector potential can be manipulated by interface engineering of the correlated itinerant ferromagnet \ce{SrRuO3} (SRO). Measurements of the AHE reveal the presence of two interface-tunable spin-polarized conduction channels. Using theoretical calculations, we show that the tunability of the AHE at SRO interfaces arises from the competition between two topologically non-trivial bands. Our results demonstrate how reconstructions at oxide interfaces can be used to control emergent electrodynamics on a nanometer-scale, opening new routes towards spintronics and topological electronics.
}\\


\noindent In topologically nontrivial band structures, electrons acquire an additional phase factor when their wavefunctions traverse a closed loop in momentum space~\cite{Berry45}. Although this concept is now commonly referred to as the Berry phase mechanism, Karplus and Luttinger already demonstrated decades earlier that the anomalous Hall effect---which is prevalent in itinerant ferromagnets---finds its origins in band topology~\cite{karplus1954hall}. In addition to the usual band dispersion contribution, electrons in an electric field $\bm{\mathcal{E}}$ acquire an anomalous velocity:

\begin{equation}
\bm{v}(\bm{k}) = \frac{1}{\hbar} \pdv{\bm{E}(\bm{k})}{\bm{k}}-\frac{e}{\hbar}\bm{\mathcal{E}} \cross \bm{b}(\bm{k}),
\end{equation}

where $E(\bm{k})$ is the dispersion relation and $\bm{b}(\bm{k})$ is the Berry curvature. The latter describes the nontrivial geometry of the band structure and acts as an effective magnetic flux~\cite{nagaosa2010anomalous}. The anomalous velocity is transverse to the electric field and gives rise to a Hall current, with a sign and magnitude that depend sensitively on the band structure topology. In systems with ferromagnetic order and sizable spin--orbit coupling (SOC), the Berry curvature is strongly enhanced near avoided band crossings which act as a source or sink of the emergent magnetic field~\cite{itoh2016weyl}. A prototypical system is the transition metal oxide SRO, a 4\textit{d} itinerant ferromagnet exhibiting an AHE that is well reproduced by first-principles calculations~\cite{fang2003anomalous}. Its anomalous Hall conductivity depends sensitively on the position of the Fermi level with respect to the avoided band crossings and on the magnetization~\cite{sinitsyn2005disorder, onoda2006intrinsic, fang2003anomalous, nagaosa2010anomalous}, forming an ideal platform to be tuned through symmetry breaking at interfaces. A suitable material for this purpose is \ce{SrIrO3} (SIO), a 5\textit{d} paramagnetic semimetal with strong atomic SOC ($\sim 0.4\;\textrm{eV}$)~\cite{moon2008dimensionality, nie2015interplay} and excellent structural compatibility with SRO. In this Letter, we investigate the AHE in ultrathin SRO films with (a)symmetric boundary conditions. We show that transport at SRO/SIO and SRO/\ce{SrTiO3} (STO) interfaces occurs through topologically non-trivial bands with opposite Berry curvature. Remarkably, in the tricolor STO/SRO/SIO system the two spin-polarized conduction channels are found to be coexisting.



\begin{figure*}[ht!]
\includegraphics[width=\linewidth]{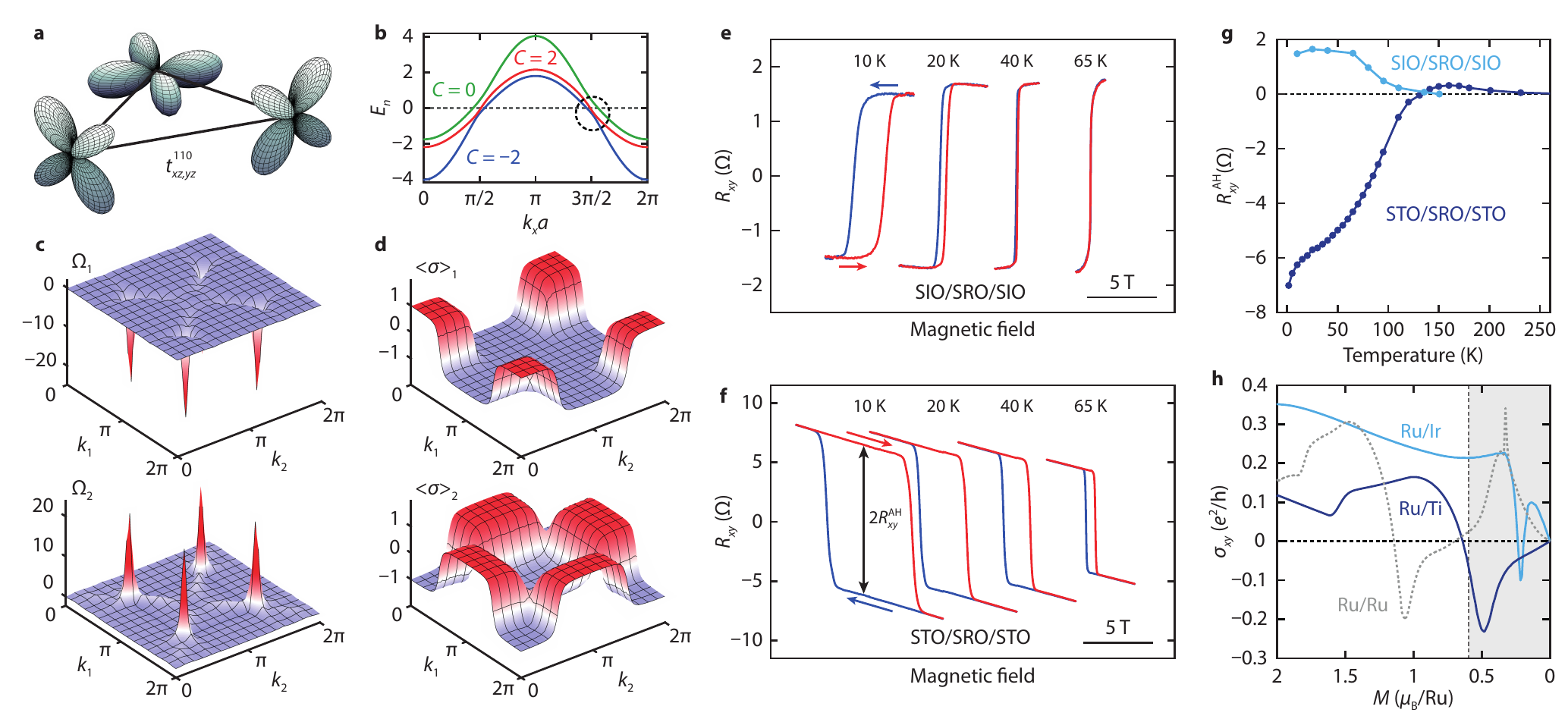}
\caption{\label{Fig1} \textbf{Anomalous Hall effect of ultrathin SRO with symmetric boundary conditions.} \textbf{a}, Next-nearest-neighbor interorbital hopping. \textbf{b}, Dispersion of \ce{Ru} $t_\textrm{2g}$ bands along $k_x = k_y$ for a representative value of the magnetization (see Supplementary Information for more details). \textbf{c}, Berry curvature associated with topologically nontrivial \ce{Ru} $t_\textrm{2g}$ bands close to the Fermi level (Chern numbers ${\cal C}=\pm 2$.). \textbf{d}, Spin polarizations $\langle\sigma^z\rangle_n$ for the corresponding bands. \textbf{e-f}, Hall resistance of symmetric SIO/SRO/SIO (\textbf{e}) and STO/SRO/STO (\textbf{f}) heterostructures as function of temperature. The curves are offset horizontally. \textbf{g}, Temperature evolution of the amplitude of the AHE ($R_{xy}^\textrm{AH}$). \textbf{h}, Evolution of the intrinsic contribution to $\sigma_\textit{xy}$ for Ru/Ti, Ru/Ir and Ru/Ru bilayers as a function of the average Ru magnetization. The dashed black line indicates the approximate saturated magnetization value of the STO/SRO/SIO determined from SQUID measurements.}
\end{figure*}
 
We first analyse theoretically the properties of ultrathin SRO starting from the Ru-based $t_\textrm{2g}$ electronic structure close to the Fermi level. Our ab-initio derived tight binding calculations show that the Berry curvature of the individual bands is strongly enhanced at avoided band crossings due to next-nearest-neighbor interorbital hopping (Fig.~\ref{Fig1}a) in the presence of SOC. We first focus on the monolayer SRO system. Its electronic structure can be arranged in two groups of 3 bands with different spin-orbital parity. Within each sector, there are two topologically nontrivial bands carrying a Chern number $C = \pm 2$, accompanied by a single, trivial band with $C = 0$, (see Fig.~\ref{Fig1}b). The ensuing Berry curvature of the nontrivial bands, which have predominantly $d_{xz}$ and $d_{yz}$ character, is shown in Fig.~\ref{Fig1}c. We find sharp peaks located at the avoided bands crossings. Since the lowest energy bands in Fig.~\ref{Fig1}b have a non-trivial Chern number, the Berry curvature contribution of each band cannot vanish and is robust to changes in the Fermi level or, in general, of the corresponding electron occupation. Their splitting and relative occupation leads to a dominance of one of the channels, including sign changes when considering the averaged Berry curvature. A complete compensation is improbable and can only accidentally occur by electronic fine tuning. SOC influences the character of the avoided crossings and causes the bands with opposite Berry curvature and $d_{xz/yz}$ orbital character to have a distinct momentum dependence of the spin polarization, with an opposite sign developing nearby the points of maximal Berry curvature accumulation, as shown in Fig.~\ref{Fig1}d.

\begin{figure*}[ht!]
\includegraphics[width=\linewidth]{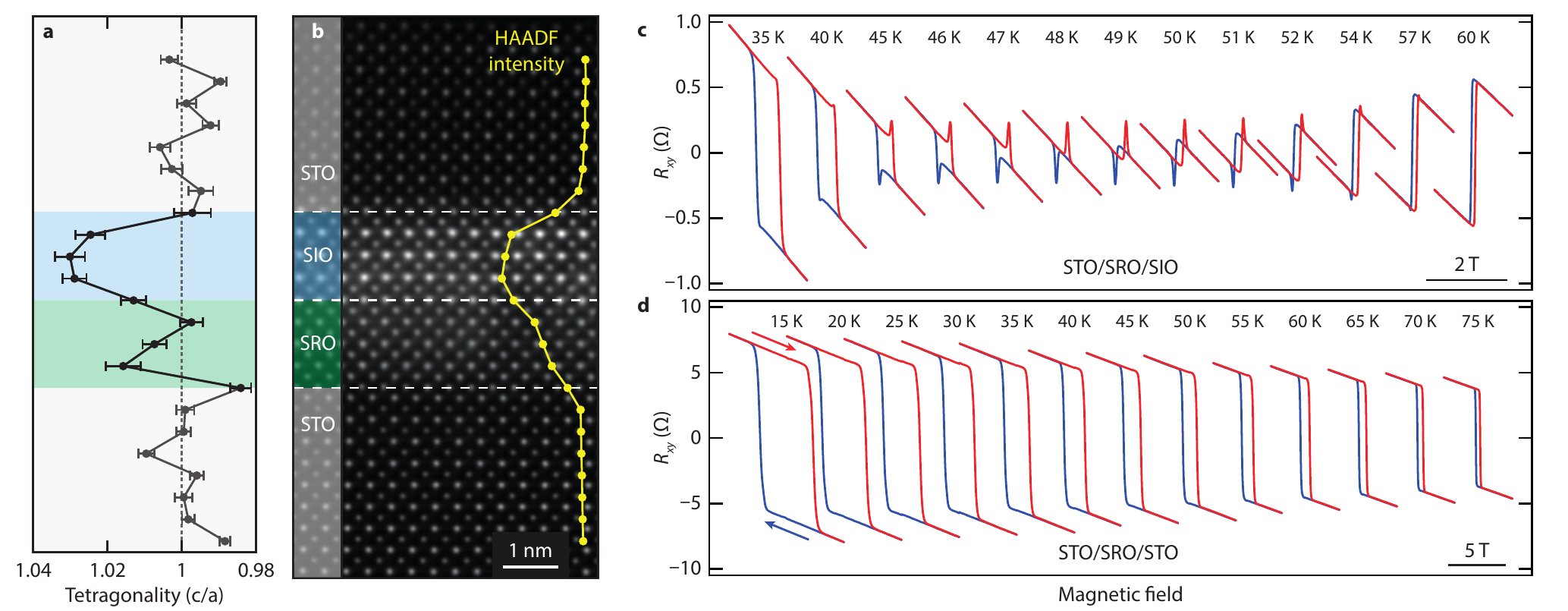}
\caption{\label{Fig2} \textbf{Anomalous Hall effect of ultrathin SRO heterostructures with asymmetric boundary conditions.} \textbf{a}, Mean tetragonality of the perovskite unit cell across the heterostructure. \textbf{b}, HAADF-STEM measurement of a STO/SRO/SIO heterostructure. \textbf{c-d}, Measured Hall resistance of (\textbf{c}) an asymmetric STO/SRO/SIO heterostructure and (\textbf{d}) a symmetric STO/SRO/STO heterostructure as function of temperature. The curves are offset horizontally.}
\end{figure*}

We now investigate SRO films with symmetric boundary conditions, shown in Fig.~\ref{Fig1}e-g. We consider heterostructures composed of STO/2 u.c.~SIO/4 u.c.~SRO/2 u.c.~SIO/10 u.c.~STO and STO/4 u.c.~SRO/10 u.c.~STO. Strikingly, we find that the sign of the AHE is opposite for SIO/SRO/SIO (Fig.~\ref{Fig1}e) and STO/SRO/STO (Fig.~\ref{Fig1}f) heterostructures. This immediately shows that symmetry breaking in ultrathin SRO directly controls the magnitude and sign of its Berry curvature. The magnitude ($R_{xy}^\textrm{AH}$) as a function of temperature is shown in Fig.~\ref{Fig1}g. While $R_{xy}^\textrm{AH}$ of the STO/SRO/STO is mainly negative and changes sign near the Curie temperature ($T_\textrm{C}$), $R_{xy}^\textrm{AH}$ of the SIO/SRO/SIO remains positive in the entire temperature range. This confirms the expectation that the occupation of the topologically active Ru $t_\textrm{2g}$ bands depends sensitively on the electronic matching at the interface. This behaviour can be qualitatively captured by modeling Ru/Ti and Ru/Ir bilayers, i.e., systems with a \ce{RuO2} monolayer coupled a \ce{TiO2} or \ce{IrO2} monolayer. As shown in Fig.~\ref{Fig1}h, for small/intermediate amplitude of the Ru magnetization the AH conductivity is negative for the Ru/Ti bilayer while it is positive for the Ru/Ir bilayer. In the former, only the Ru $d_{xz/yz}$ contribute since the STO is electronically inert, while for the latter, the intrinsic Berry curvature sign competition of the Ru topological bands is modified through the hybridization of the Ir/Ru $d_{xz/yz}$  orbitals, and interfacial magnetic canting/reconstructions (see Supplementary Information).


To study the effect of asymmetric boundary conditions, we now investigate the tricolor STO/SRO/SIO system (Fig.~\ref{Fig2}). Given the different trends observed in the symmetric systems, we expect competition in the total $R_{xy}^\textrm{AH}$ in this case. The atomic arrangement at the interfaces is investigated by high-angle annular dark-field scanning transmission electron microscopy (HAADF-STEM) imaging (Fig~\ref{Fig2}b). Chemical analysis by electron energy loss spectroscopy (EELS) shows that the interfaces are atomically sharp and that the thicknesses of both the SRO and SIO layers are 4 u.c.~as designed (see Supplementary Information). After quantifying atomic column positions in the HAADF-STEM image using StatSTEM~\cite{de2016statstem}, a detailed analysis of the atomic positions shows that octahedral tilts are suppressed and both the SRO and SIO are tetragonal rather than orthorhombic as in their bulk form. In addition, we find that the tetragonality ($c/a$) of the unit cell varies strongly across the SRO and SIO layers (Fig.~\ref{Fig2}a). Since the magnetic anisotropy of SRO is known to be very sensitive to strain and tetragonality~\cite{gan1998direct, jung2004magnetic, kan2013epitaxial}, this affects the easy axis direction of the different SRO layers and hence the local magnetization of the \ce{Ru} ions. This is confirmed by SQUID measurements, included in the Supplementary Information, which show that the STO/SRO/SIO has a larger in-plane magnetization than STO/SRO/STO. This indicates that the magnetization of the SRO layers near the SIO interface is canted, which is consistent with the reduction of $c/a$ close to the SIO interface.

The AHE of the STO/SRO/SIO is shown in Fig.~\ref{Fig2}c. With increasing temperature, the AHE changes sign at the reversal temperature $T_\textrm{R} = 48\;\textrm{K}$ and peaks appear to be superimposed on the Hall effect, slightly above and below the coercive field ($B_c$). This is in stark contrast with the AHE of an STO/SRO/STO heterostructure (Fig.~\ref{Fig2}d), where the magnitude ($R_{xy}^\textrm{AH}$) decreases with increasing temperature. The peaks superimposed on the Hall effect are present between 35 and 58 K and reach their maximum amplitude at $T_\textrm{R}$, i.e., when $R_{xy}^\textrm{AH}$ appears to be zero. This strongly suggests that their occurrence is intrinsically linked to the sign reversal of the AHE. In the following, we will show that the AHE can be modeled by the superposition of two anomalous Hall components with opposite sign, arising from two spin-polarized conduction channels. We attribute this to a modified band occupation of the SRO layers near the STO and SIO interfaces, giving rise to two conduction channels with a different $R_{xy}^\textrm{AH}(T)$ dependence. Within this picture, $T_\textrm{R}$ of the SRO layers near the SIO and STO interfaces no longer coincide, resulting in a temperature window in which $R_{xy}^\textrm{AH}$ of the two channels is of opposite sign. This scenario is supported by the observation of a reduced tetragonality at the SRO/SIO interface by HAADF-STEM (see above) and related canted ferromagnetic moment (see Supplementary Information). It is also supported by our theoretical calculations of 2 u.c.~STO/4 u.c.~SRO/2 u.c.~SIO heterostructures, which show that the system can be approximated by two effective electronic channels. This situation arises due to the competition between the intrinsic ferromagnetism of the SRO and the magnetic reconstructions at the interface (see Supplementary Information).

\begin{figure*}[ht!]
\includegraphics[width=\linewidth]{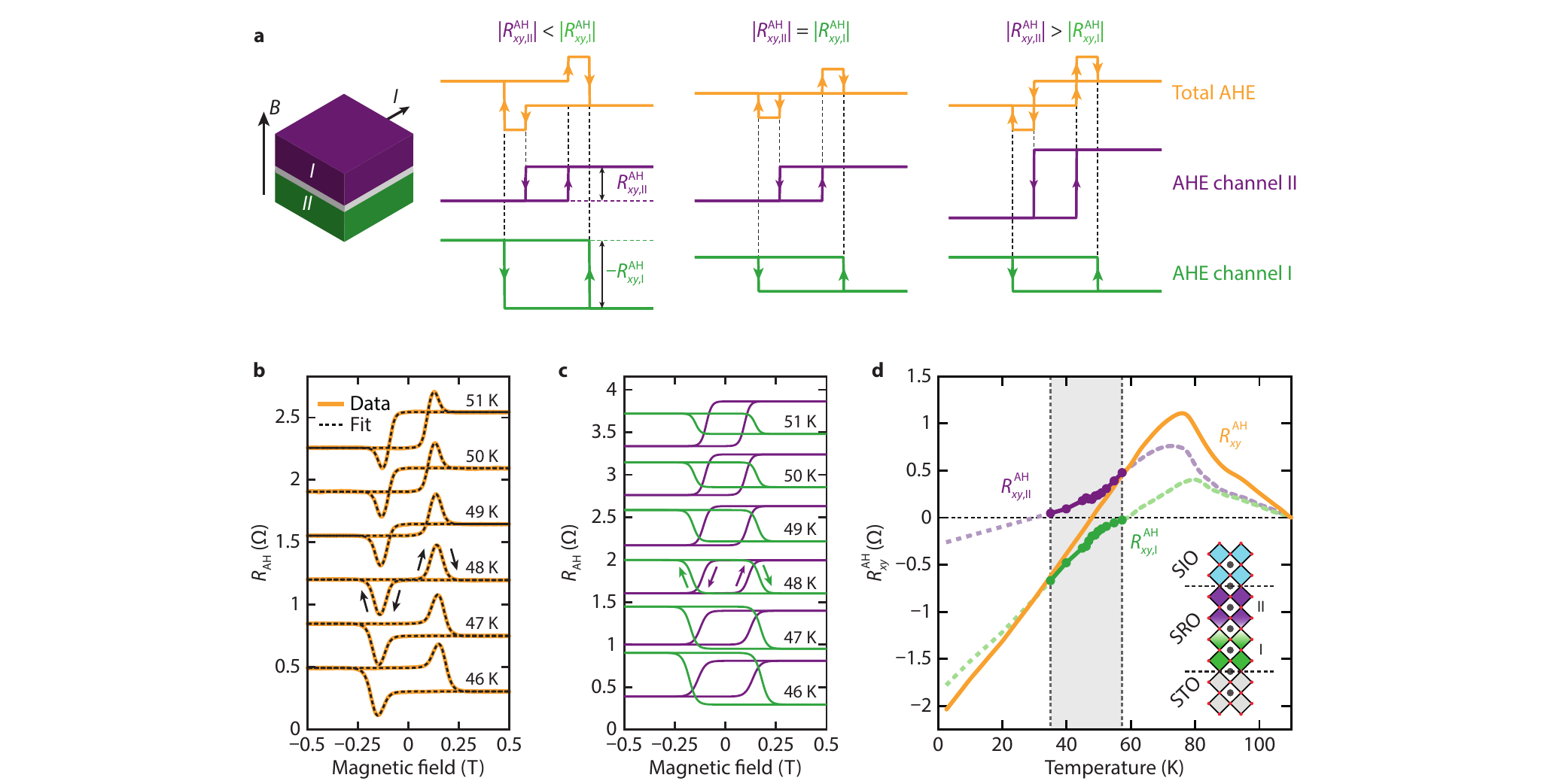}
\caption{\label{Fig3}\textbf{Presence of two anomalous Hall channels.} \textbf{a}, Addition of the AHE for two decoupled ferromagnetic layers with opposite sign of $R_{xy}^\textrm{AH}$. \textbf{b}, $R_\textrm{AH}$ as function of temperature. The black dashed lines are fits to the data and the curves are offset vertically. \textbf{c}, The two anomalous Hall components that add up to the total $R_\textrm{AH}$ curves in \textbf{b}. \textbf{d}, Total $R_{xy}^\textrm{AH}$ and the extracted $R_{xy}^\textrm{AH}$ from the two anomalous Hall components. The dashed lines illustrate a possible temperature dependence of $R_{xy}^\textrm{AH}$, and the inset sketches a potential spatial profile.}
\end{figure*}


To illustrate the total AHE in this case, we consider a heterostructure with two independent anomalous Hall channels (labeled I and II) with $R_{xy}^\textrm{AH}$ of opposite sign and $B_\textrm{c,II}<B_\textrm{c,I}$. This situation is sketched in Fig.~\ref{Fig3}a for three cases: $|R_{xy,\textrm{II}}^\textrm{AH}|<|R_{xy,I}^\textrm{AH}|$ (left), $|R_{xy,\textrm{II}}^\textrm{AH}|=|R_{xy,\textrm{I}}^\textrm{AH,}|$ (middle), and $|R_{xy,\textrm{II}}^\textrm{AH}|>|R_{xy,\textrm{I}}^\textrm{AH}|$ (right). When a current $I$ is applied in the plane of the heterostructure and the magnetic field is varied in the range $[0,B,-B,0]$, the total AHE is given by the sum of the AHE of the two layers. Depending on their relative magnitudes, three different behaviors can be discerned for $B_\textrm{c,II}<B_\textrm{c,I}$. This concept was first introduced in 1981~\cite{kobayashi1981magnetization} and forms the basis for a device called the extraordinary Hall balance~\cite{zhang2013extraordinary, zhang2016magneto}.

\begin{figure*}[ht!]
\includegraphics[width=\linewidth]{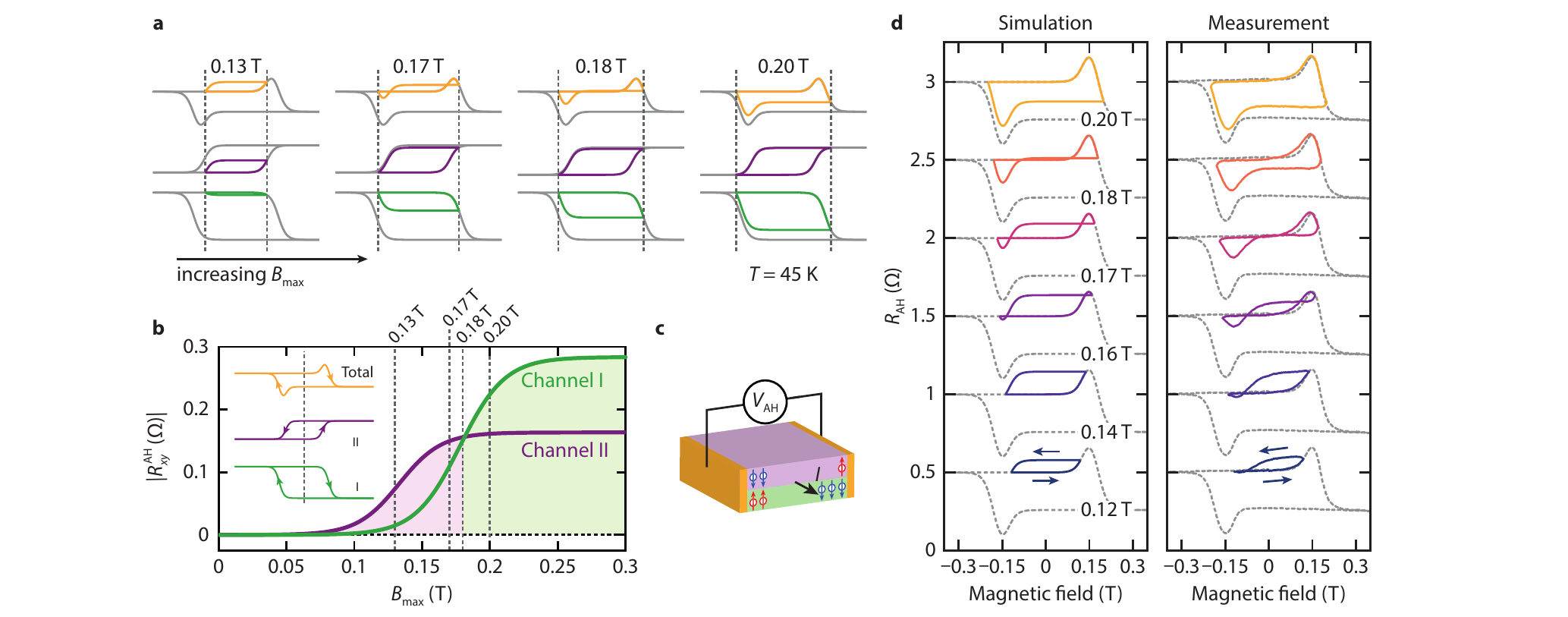}
\caption{\label{Fig4}\textbf{Control of the spin polarization of the two channels.} \textbf{a}, Simulated $R_\textrm{AH}$ curves for different values of $B_\textrm{max}$. \textbf{b}, $|R_{xy}^\textrm{AH}|$ of the separate AH components as a function of $B_\textrm{max}$. \textbf{c}, Schematic showing the accumulation of spin-polarized electrons on opposite sides of the channel. \textbf{d}, Simulated (left) and measured (right) $R_\textrm{AH}$ curves. The curves are offset vertically.}
\end{figure*}

The measured AHE of the STO/SRO/SIO in Fig.~\ref{Fig2}c bears a striking resemblance to the curves in Fig.~\ref{Fig3}a. The ordinary Hall component has been subtracted and the remaining anomalous Hall component ($R_\textrm{AH}$) is presented in Fig.~\ref{Fig3}b. As the temperature is increased from $\SI{46}{K}$ towards $\SI{51}{K}$, the behavior of the total AHE evolves from the leftmost scenario in Fig.~\ref{Fig3}a to the rightmost scenario, with the middle scenario emerging at $T_\textrm{R}=\SI{48}{K}$. This is consistent with two anomalous Hall contributions of opposite sign, each with a slightly different temperature dependence. We investigate this further by considering a phenomenological model of the AH data with $R_\textrm{AH} = R_{xy,\textrm{I}}^\textrm{AH}\tanh(\omega_\textrm{I}(B-B_\textrm{c,I})) + R_{xy,\textrm{II}}^\textrm{AH}\tanh(\omega_\textrm{II}(B-B_\textrm{c,II}))$, where $\omega$ is a parameter describing the slope at $B_\textrm{c}$. An excellent agreement is obtained between this model (dashed black lines in Fig.~\ref{Fig3}b) and the data, enabling us to extract the individual AH components as a function of temperature (Fig.~\ref{Fig3}c). The corresponding $R_{xy}^\textrm{AH}$ values are shown in Fig.~\ref{Fig3}d; both components show a smooth evolution in temperature, with one disappearing above $\SI{58}{K}$ (green) and the other below $\SI{35}{K}$ (purple). At $\SI{48}{K}$ the two components are equal, leading to the fully compensated case. 

In Fig.~\ref{Fig3}d we illustrate a possible dependence of $R_{xy}^\textrm{AH}(T)$ (dashed lines) at higher and lower temperatures, which suggests that $R_{xy}^\textrm{AH}$ and $B_\textrm{c}$ of the two channels follow a qualitatively similar temperature dependence, shifted by $\SI{23}{K}$. This implies that, for $T < \SI{35}{K}$ and $T > \SI{58}{K}$, $R_{xy}^\textrm{AH}$ of the two channels are of the same sign or the positive contribution is below the detection limit of our experiment, rendering the total AHE indistinguishable from that of a single spin-polarized channel (see Supplementary Information). However, it should be noted that any two curves that add up to the total $R_{xy}^\textrm{AH}$ are in principle possible. The different $B_\textrm{c}$ and $T_\textrm{R}$ of the two channels can be attributed to the different anisotropy and magnitude of the interfacial \ce{Ru} magnetization.


The presence of two channels with different switching field $B_\textrm{c}$ provides the possibility to control their relative spin polarizations by choosing the magnetic field span interval $B_\textrm{max}$ appropriately. By varying $B$ in the range $[0,B_\textrm{max},-B_\textrm{max},0]$, the polarization of one channel can be (partially) switched while the other is less affected. This is illustrated in Fig.~\ref{Fig4}a for different values of $B_\textrm{max}$. The simulations are performed by traversing the individual $R_\textrm{AH}$ curves (determined from the data at $\SI{45}{K}$) up to $B_\textrm{max}$ and summing them to obtain the total $R_\textrm{AH}$. It follows that increasing $B_\textrm{max}$ changes the relative magnitudes of $R_{xy,I}^\textrm{AH}$ and $R_{xy,II}^\textrm{AH}$, replicating the temperature evolution of the AHE. This is summarized in the diagram in Fig.~\ref{Fig4}b, where the magnitudes of $R_{xy,I}^\textrm{AH}$ and $R_{xy,II}^\textrm{AH}$ are plotted against $B_\textrm{max}$. The crossing of the curves at $0.18\;\textrm{T}$ constitutes the compensation point where $|R_{xy,I}^\textrm{AH}| = |R_{xy,II}^\textrm{AH}|$ and the height of the apparent peaks is maximum.

The behavior predicted by the simulations is indeed found in the measurements, which is showcased in Fig.~\ref{Fig4}d. A difference between the experimental data and the simulations is found at low $B_\textrm{max}$, where an asymmetry is observed in the measured $R_\textrm{AH}$ curves. This can be attributed to a degree of interlayer coupling, quantifiable in the order of tens of mT, which reduces the field required to restore the original magnetic state. Given that the channels have the same spin polarization and carrier type, the opposite anomalous Hall voltages indicate that spin-polarized electrons are accumulated on opposite sides of the two channels as sketched in Fig.~\ref{Fig4}c for $B = -0.5\;\textrm{T}$. Since the two interfaces are separated by a few u.c., overlap between their respective wavefunctions is indeed possible. However, the much larger in-plane hopping parameters found by density functional theory (DFT) (see Supplementary Information) support the effective model of two spin-polarized channels. Within this picture, the system is brought to a metastable state at $B_\textrm{max}$, which microscopically can be viewed as opposite majority spins being accumulated above each other. Such a configuration is energetically unfavorable and the system tends to restore the initial spin distribution, causing the asymmetry observed in the data.


Finally, we compare our results to recent work on SRO thin films and interfaces, where similar anomalous Hall characteristics were observed and attributed to the topological Hall effect due to a skyrmion phase~\cite{matsuno2016interface, pang2017spin, ohuchi2018electric}. Within this picture, the topological Hall effect should be enhanced in the SIO/SRO/SIO case due to Dzyaloshinskii-Moriya interaction at both interfaces. Instead, we find that this system displays a regular anomalous Hall effect due to the imposition of symmetric boundary conditions. Additionally, in this scenario the skyrmion phase is present around the temperature at which the AHE changes sign, which would be highly coincidental. However, two anomalous Hall components with a shifted $R_{xy}^\textrm{AH}(T)$ dependence naturally produces this temperature window around $T_\textrm{R}$, as there the AHE is of opposite sign. We also show in the Supplementary Information that considering an anomalous and topological Hall contribution results in an unphysical discontinuity of the coercive field, which is not present in our model. Finally, our model can quantitatively describe the behavior of $R_{xy}(B)$, and correctly predicts the behavior when sweeping the magnetic field to values below saturation. Recent work has shown that \ce{Ru} vacancies can also affect the AHE, but the studied system also presents an asymmetry in the form of dissimilar interfaces and a gradient of octahedral rotations~\cite{kan2018defect} which can explain the observed characteristics.

The atomic-scale control of spin and charge accumulation through Berry phase engineering opens new avenues for spintronic devices and topological electronics. In this respect, transition metal oxides are an ideal platform owing to a delicate interplay between spin, charge and lattice degrees of freedom. Our results establish that oxide interfaces host tunable topological phenomena, thereby providing new perspectives in the field of complex oxides.


\section*{Methods}

\subsection*{Sample fabrication}

SRO/STO, SRO/SIO/STO, and SIO/SRO/SIO/STO heterostructures were prepared by pulsed laser deposition on \ce{TiO2}-terminated STO(001) substrates (CrysTec GmbH). The films were deposited at $600^\circ\textrm{C}$ in an oxygen pressure of $0.1\;\textrm{mbar}$. The laser fluence was $1.2\;\textrm{J}/\textrm{cm}^2$ and the repetition rate was $1\;\textrm{Hz}$. To refill possible oxygen vacancies formed during the growth, the samples were annealed at $550^\circ\textrm{C}$ in an oxygen pressure of $300\;\textrm{mbar}$ and cooled down in the same pressure at a rate of $20^\circ\textrm{C}$/min. The growth was monitored by reflection high-energy electron diffraction (RHEED), indicating a layer-by-layer growth mode for the three films. 

\subsection*{Structural \& magnetotransport characterization}

Atomic scale characterization of the lattice structure was performed on an aberration corrected STEM. The FEI Titan 80-300 microscope was operated at $300\;\textrm{kV}$ and the samples were prepared in a vacuum transfer box and analyzed in a Gatan Vacuum transfer holder to avoid any influence of air on the film~\cite{grieten2017reclaiming, conings2017structure}. Collection angles for HAADF imaging, ABF imaging and EELS were $44$-$190\;\textrm{mrad}$, $8$-$17\;\textrm{mrad}$ and $47\;\textrm{mrad}$, respectively. The interfaces are atomically sharp and the STEM-EELS measurements show that there is no diffusion of Ru and Ti, whereas there is a slight diffusion of Ir into the top \ce{RuO2} layer. The heterostructures were further investigated by synchrotron X-ray diffraction measurements and scanning tunneling microscopy (see Supplementary Information). Hall bars were patterned by e-beam lithography, and the heterostructure was contacted by Ar etching and \textit{in-situ} deposition of Pd and Au, resulting in low-resistance Ohmic contacts. An STO cap layer was used to impose symmetric boundary conditions and prevent degradation of the SIO layer~\cite{groenendijk2016epitaxial, groenendijk2017spin}. Transport measurements were performed in a He flow cryostat with a $10\;\textrm{T}$ superconducting magnet and a base temperature of $1.5\;\textrm{K}$. Measurements in current-bias configuration were performed using lock-in amplifiers and custom-made low noise current sources and voltage amplifiers.

\subsection*{DFT calculations}

First-principles DFT calculations were performed using the VASP~\cite{kresse1996efficiency} package based on plane wave basis set and projector augmented wave method~\cite{kresse1999ultrasoft}. A plane-wave energy cut-off of $500\;\textrm{eV}$ was used. For the treatment of exchange-correlation, the LSDA (local spin density approximation) with the Perdew-Zunger~\cite{perdew1981self} parametrization of the Ceperly-Alder data~\cite{ceperley1980ground} for the exchange-correlation functional was considered. The choice of LSDA exchange functional is suggested by a recent paper~\cite{etz2012indications}, where it was shown that LSDA is a better approximation than the Generalized Gradient Approximation for bulk SRO and its heterostructures~\cite{roy2015interface,autieri2016antiferromagnetic}. In our simulations, the STO/SRO/SIO heterostructure was constructed using a lateral supercell of $\sqrt{2}a \times \sqrt{2}a$, while the phases without rotations were contracted using a lateral supercell of $a \times a$. The in-plane lattice parameter was fixed to that of the STO substrate, while for the out-of-plane lattice parameters we used the experimental values of the single unit cell of SRO and SIO. The hopping parameters were estimated from the electronic structure of the non-magnetic SRO/SIO and SRO/STO interfaces without Coulomb repulsion. After obtaining the Bloch wave functions from DFT, the maximally localized Wannier functions~\cite{marzari1997maximally, souza2001maximally} were constructed using the WANNIER90 code~\cite{mostofi2008wannier90}. Starting from an initial projection of atomic $d$-basis functions belonging to the $t_{2g}$ manifold and centered on metal sites, we obtained the $t_{2g}$-like Wannier functions. To extract the hopping parameters from the electronic bands at low energies, we used the Slater-Koster interpolation as implemented in WANNIER90. This approach is applied to determine the real space Hamiltonian matrix elements in the $t_\textrm{2g}$-like Wannier function basis for the SRO/SIO and SRO/STO interfaces.

\subsection*{Data availability}
\noindent The data that support the findings of this study are available from the corresponding authors upon reasonable request.

\section{Acknowledgements}
\noindent The authors thank Y.~Blanter, G.~Koster, M.~Golden, A.~M.~R.~V.~L.~Monteiro, I.~Lindfors-Vrejoiu, M.~Gabay, J.~M.~Triscone, S.~Gariglio, and M.~Kawasaki for discussions. This work was supported by The Netherlands Organisation for Scientific Research (NWO/OCW) as part of the Frontiers of Nanoscience program (NanoFront), by the Research Foundation Flanders (FWO, Belgium), and by the European Research Council under the European Union’s Horizon 2020 programme/ERC Grant Agreements No.~[677458], No.~[770887] and No.~[731473] (Quantox of QuantERA ERA-NET Cofund in Quantum Technologies). C.A.~and S.P.~acknowledge financial support from Fondazione Cariplo via the project Magister (Project No. 2013-0726) and from CNR-SPIN via the Seed Project ``CAMEO''. C.A.~acknowledges the CINECA award under the ISCRA initiative IsC54 ``CAMEO'' Grant, for the availability of high performance computing resources and support. W.B.~acknowledges support by Narodowe Centrum Nauki (NCN, National Science Centre, Poland) Project No.~2016/23/B/ST3/00839. The work is supported by the Foundation for Polish Science through the IRA Programme co-financed by the European Union within SG OP. N.G.~and J.V.~acknowledge support from the GOA project ``Solarpaint'' of the University of Antwerp. The Qu-Ant-EM microscope was partly funded by the Hercules fund from the Flemish Government.

\section{Author contributions}
\noindent D.J.G.~deposited and characterized the epitaxial heterostructures, fabricated the Hall bars and performed the (magneto)transport measurements with help from T.C.T. D.J.G.~and T.C.T.~analysed the transport data and simulated the anomalous Hall curves. N.G., K.H.W.~v.d.~B, S.~v.~A.~and J.V.~measured and analysed the HAADF-STEM and STEM-EELS data. C.A.~performed the DFT calculations, took part in the analysis of the theoretical data and supervised the theoretical part. S.P.~and P.B.~took part in the analysis of theoretical data. W.B. performed the symmetry analysis of the effective model and found the spin-orbital parity resolved Chern numbers. M.C. supervised the theoretical work related to the topological aspects of the effective model. W.B. and M.C. calculated the Berry curvature and both took part in the analysis of theoretical data. A.D.C.~supervised the overall project. D.J.G., T.C.T.~and A.D.C.~wrote the manuscript with input from all authors.

\section{Additional information}
Correspondence and requests for materials should be addressed to D.J.G.~or A.D.C.

\section{Competing financial interests}
\noindent The authors declare no competing financial interests.


\bibliography{References}

\clearpage


\section*{Supplementary material}

# SUPPLEMENTARY INFORMATION FOR:
# BERRY PHASE ENGINEERING AT OXIDE INTERFACES

Dirk J. Groenendijk,[1, *] Carmine Autieri,[2, †] Thierry C. van Thiel,[1, †] Wojciech Brzezicki,[2, 3] Nicolas Gauquelin,[4] Paolo Barone,[2] Karel H. W. van den Bos,[4] Sandra van Aert,[4] Johan Verbeeck,[4] Alessio Filippetti,[5, 6] Silvia Picozzi,[2] Mario Cuoco,[2, 7] and Andrea D. Caviglia[1, ‡]

[1]*Kavli Institute of Nanoscience, Delft University of Technology,*
*P.O. Box 5046, 2600 GA Delft, Netherlands*
[2]*Consiglio Nazionale delle Ricerche CNR-SPIN, Italy*
[3]*International Research Centre MagTop at Institute of Physics,*
*Polish Academy of Sciences, Aleja Lotników 32/46, PL-02668 Warsaw, Poland*
[4]*Electron Microscopy for Materials Science (EMAT),*
*University of Antwerp, 2020 Antwerp, Belgium*
[5]*Dipartimento di Fisica, Università di Cagliari, Cagliari, Monserrato 09042-I, Italy*
[6]*CNR-IOM, Istituto Officina dei Materiali,*
*Cittadella Universitaria, Cagliari, Monserrato 09042-I, Italy*
[7]*Dipartimento di Fisica "E. R. Caianiello" Universitá degli Studi di Salerno, 84084 Fisciano, Italy*

## CONTENTS

---
* d.j.groenendijk@tudelft.nl
† equal contribution
‡ a.caviglia@tudelft.nl



# I. STRUCTURAL AND MAGNETIC CHARACTERIZATION

## A. Crystalline quality and magnetic anisotropy

In Fig. S1 we present the structural and magnetic characterization of our films using various methods. Figure S1a shows the RHEED intensity during the growth of a 4/4/10 u.c. SRO/SIO/STO heterostructure on a TiO$_2$-terminated STO(001) substrate. The clear intensity oscillations indicate that all three layers grow in layer-by-layer mode, and the strong initial oscillation during the SRO growth can be attributed to a change in surface termination from TiO$_2$ to SrO.

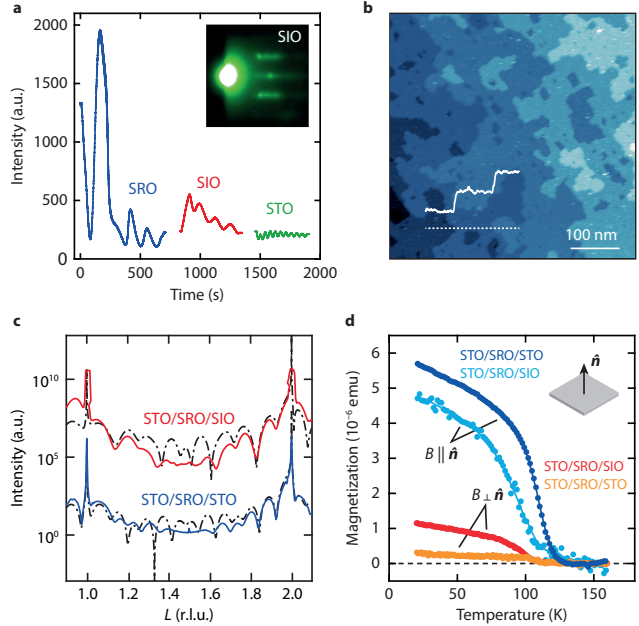

FIG. S1: **Structural and magnetic characterization. a**, RHEED intensity oscillations during the growth of SRO, SIO, and STO. Inset: diffraction pattern after SIO growth. **b**, STM image of the surface of an STO/SIO/SRO heterostructure. **c**, XRD measurements of STO/SRO/STO and STO/SRO/SIO heterostructures. The black dashed lines are simulations of the diffracted intensity. **e**, In- and out-of-plane $M(T)$ of STO/SRO/STO and STO/SRO/SIO. The data was acquired during field-cooling with an applied field of 50 mT.

Figure S1b and c show an STM topographic map and synchrotron diffraction measurements, respectively. The atomically flat surface and clear Laue oscillations indicate that the heterostructures are of high crystalline quality and have sharp interfaces. The magnetization $M$ of an STO/SRO/STO and an STO/SRO/SIO heterostructure (with 4 u.c. SRO and 2 u.c. SIO) are measured by SQUID magnetometry. Figure S1d shows the in- and out-of-plane magnetization of the two samples. Both films have a predominantly out-of-plane magnetization with values reaching approximately 0.6 and 0.8 $\mu_B$/Ru at 20 K for STO/SRO/STO and STO/SRO/SIO, respectively. Compared to the STO/SRO/STO, the out-of-plane magnetization of the STO/SRO/SIO is slightly reduced and an in-plane component appears.

## B. Compositional analysis by STEM-EELS

The atomic structure of each interface was investigated using high-angle annular dark-field scanning transmission electron microscopy (HAADF-STEM) combined with electron energy loss spectroscopy (EELS). An ADF measurement and EELS chemical maps at the Ti-$L_{2,3}$, Ru-$M_{4,5}$, Sr-$L_{2,3}$ and Ir-$L_{4,5}$ edges are shown in the first 5 panels of Fig. S2. The Ir M edge is at high energy (2 keV) and very close to the Sr edge, making it difficult to discern.

A combined EELS chemical map (panel 6 of Fig. S2) is obtained from the normalized integrated intensities of the aforementioned edges. The color code in this color composite is Sr (red), Ti (yellow), Ir (blue) and Ru (green). A profile of the normalized integrated intensities along the growth direction is shown in the rightmost panel. Since the A-site ion is shared throughout the heterostructure, the interfaces are formed at the B-sites (only the B-site profile is displayed). The Ti/Ru and Ti/Ir interfaces are atomically sharp, and only a single B-site layer displays intermixing. At the Ru/Ir interface, a slight diffusion of Ir into the topmost $RuO_2$ layer is observed, which corresponds to an Ir content of approximately $20 \pm 10\%$.

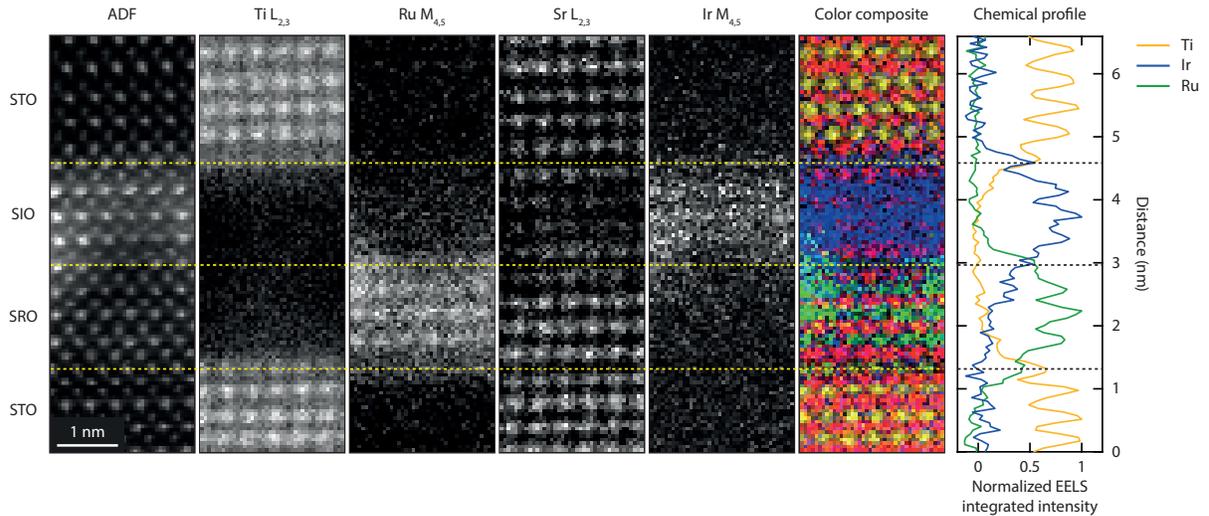

FIG. S2: **Atomic characterization of the heterostructure.** From left to right: ADF atomic Z-contrast image, with simultaneously acquired EELS maps of the Ti-$L_{2,3}$, Ru-$M_{4,5}$, Sr-$L_{2,3}$ and Ir-$M_{4,5}$ edges for an SRO/SIO heterostructure. All atomic maps are overlapped to form the color composite shown on the next panel with Sr (red), Ti (yellow), Ir (blue) and Ru (green). The corresponding normalized intensity profile of the B-sites is shown in the next panel.

### C. Determination of lattice parameters and octahedral tilt angles

The in- and out-of-plane lattice parameters of the STO/SRO/SIO heterostructure (used to calculate the tetragonality in Fig. 2a of the main text) were determined by extracting the atomic positions from the HAADF-STEM image. StatSTEM was used to model the image as a superposition of Gaussian peaks using statistical parameter estimation theory [1], which takes into account the overlap of neighbouring intensities. The tetragonality is then determined by dividing the out-of-plane ($c$) over the in-plane ($a$) lattice parameter of each unit cell. These measurements are averaged row-by-row and the corresponding errors are calculated.

The ABF image of the heterostructure is shown in Fig. S3a. Analysis of the oxygen positions show that there are no octahedral tilts across the heterostructure and that the films are in a tetragonal state (see Fig. S3b). The octahedral tilt was determined in the same manner as in Refs. [2, 3]. The final profile (Fig. S3b) shows the averaged octahedral tilt row-by-row. The presented errors are the errors on these mean values.

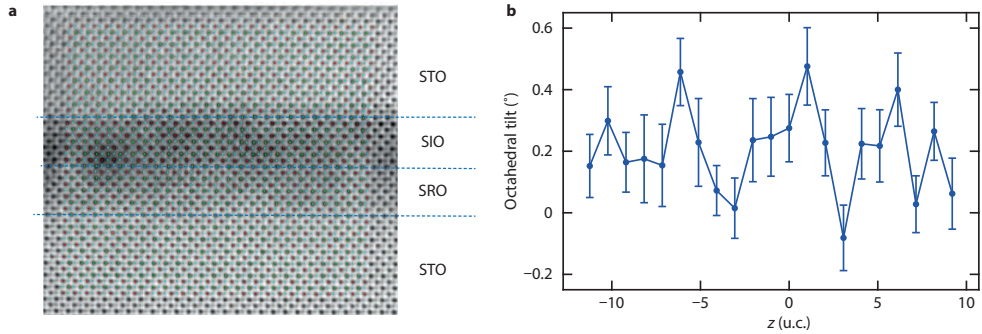

FIG. S3: **Analysis of octahedral tilts. a,** ABF image of the STO/SRO/SIO heterostructure. **b,** Octahedral tilt angles determined from the oxygen positions.

## II.  MAGNETOTRANSPORT

### A.  Anomalous Hall effect of (a)symmetric heterostructures

Figure S4 shows the AHE of Fig. 1b, c in a wider temperature range. Careful analysis of the hysteresis loops enables the extraction of the loop height ($R_{xy}^{AH}$) and the coercive field ($B_c$) as a function of temperature.

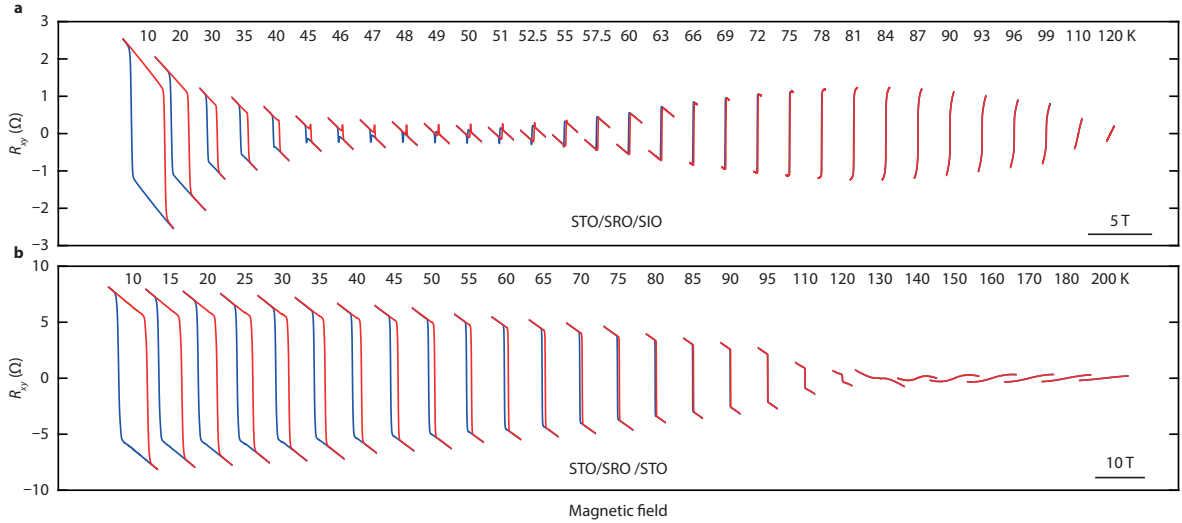

FIG. S4:  **Anomalous Hall effect.** AHE for a range of temperatures of **a**, an STO/SRO/SIO heterostructure and **b**, an STO/SRO/STO heterostructure, both with a 10 u.c. STO cap layer. The AHE of the STO/SRO/STO shows a sign reversal close to $T_C$, in agreement with previous measurements on bulk and thin-film SRO.

Figure S5a shows $R_{xy}^{AH}$ of both samples in the range $1.5\,\mathrm{K} \leq T \leq 120\,\mathrm{K}$. When there is a clear hysteresis loop, $R_{xy}^{AH}$ can be determined from the difference between the forward and backward traces at $B = 0\,\mathrm{T}$. At higher temperatures, the switching of the magnetization becomes less abrupt, making it difficult to separate the ordinary and anomalous Hall components and accurately determine $R_{xy}^{AH}$. Therefore, for the STO/SRO/SIO heterostructure the $R_{xy}^{AH}$ values were determined with an associated uncertainty range for $T \geq 57.5\,\mathrm{K}$ (see Figs. S5b and c). Due to the higher $T_C$ and larger coercive field of the STO/SRO/STO, an estimation of the uncertainty was only required for $T \geq 110\,\mathrm{K}$.

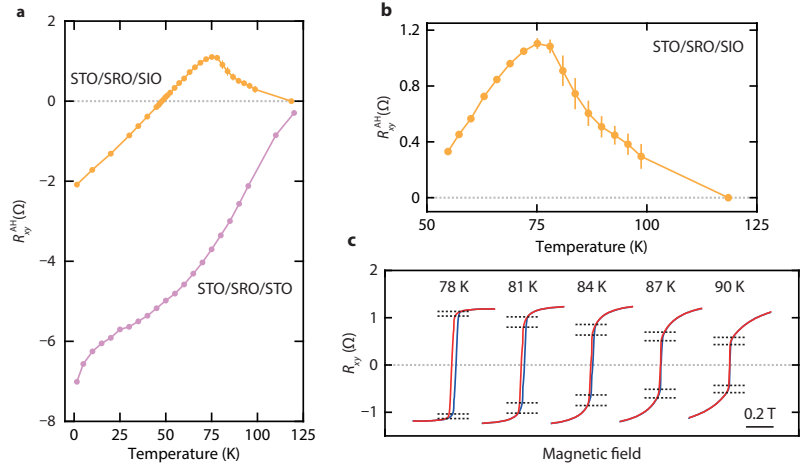

FIG. S5: **Determination of $R_{xy}^{AH}$. a,** $R_{xy}^{AH}$ as a function of temperature of STO/SRO/SRO and STO/SRO/SIO heterostructures. **b,** Enlarged portion of **a**. **c,** Illustration of the uncertainty in $R_{xy}^{AH}$.

### B. Determination of the coercive field

$B_c$ can be determined from the AHE by finding the first zero-crossing of $R_{xy}$ for $B > 0$. This procedure is shown in Fig. S6a for the AHE of the STO/SRO/SIO heterostructure around $T_R$. The extracted $B_c$ values are plotted versus temperature in Fig. S6b (left), which reveals an unphysical discontinuity occurring at $T_R$. The coercive fields determined from the fits with two anomalous Hall components (right) show a continuous behaviour as a function of temperature, which lends strong support to the presence of two anomalous Hall channels. Notably, the carrier density (Fig. S6c) determined from the ordinary Hall component does not show an anomaly around $T_R$.

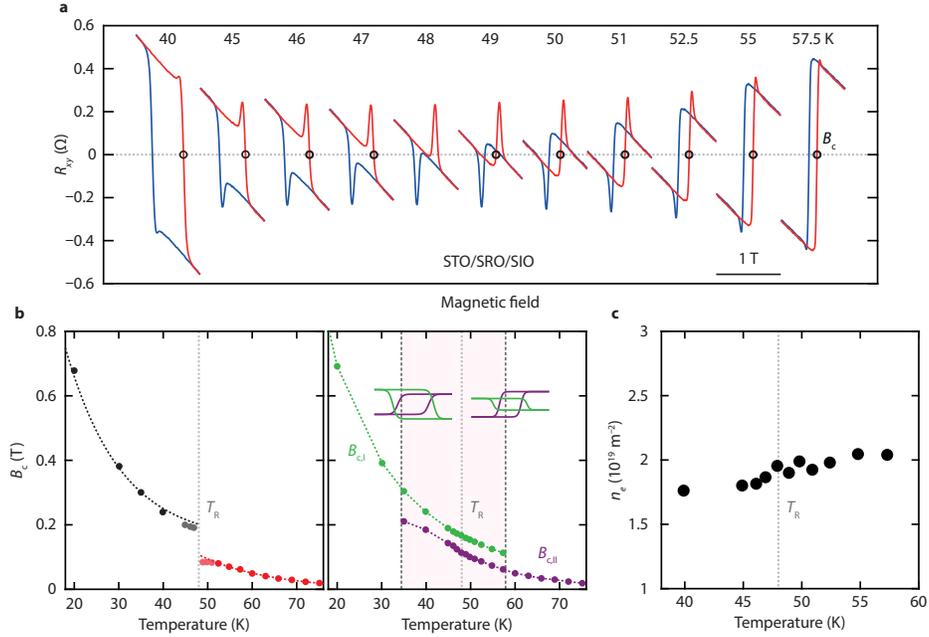

FIG. S6: **Determination of $B_c$. a,** AHE around the sign reversal. The circles indicate the determined $B_c$. **b,** Left: $B_c$ as a function of temperature. At $T_R$, $B_c$ displays a discontinuity. Right: $B_c$ determined from the fits with two anomalous Hall components. **c,** Carrier density ($n_e$) as a function of temperature, determined from the high-field slopes in **a**.

## C. Extrapolation of the anomalous Hall components in temperature

In Fig. S7 we investigate the anomalous Hall effect outside of the temperature window where $R_{xy}^{\mathrm{AH}}$ is of opposite sign ($35\,\mathrm{K} \leq T < 58\,\mathrm{K}$). Based on the extrapolation of the data in Fig S7a, we expect $R_{xy}^{\mathrm{AH}}$ of both channels to be negative for $T < 35\,\mathrm{K}$ and positive for $T > 58\,\mathrm{K}$. We elucidate this hypothesis in Fig. S7b by summing the extrapolated loops (green and purple), assuming extrapolated values for $B_{\mathrm{c}}$ at 30 and 60 K from Fig. S6a. The result of the summation (orange) is shown in the panel below, and is compared to a single-loop fit of the data (black lines) in Figs. S7c and d. Due to (i) the slight difference in $B_{\mathrm{c}}$, (ii) the considerable slope at $B_{\mathrm{c}}$ and (iii) the fact that $R_{xy}^{\mathrm{AH}}$ of one channel is always much smaller than the other, the difference is extremely subtle. This implies that, outside the discussed temperature window, it is nearly impossible to distinguish the two components from the AHE measurements.

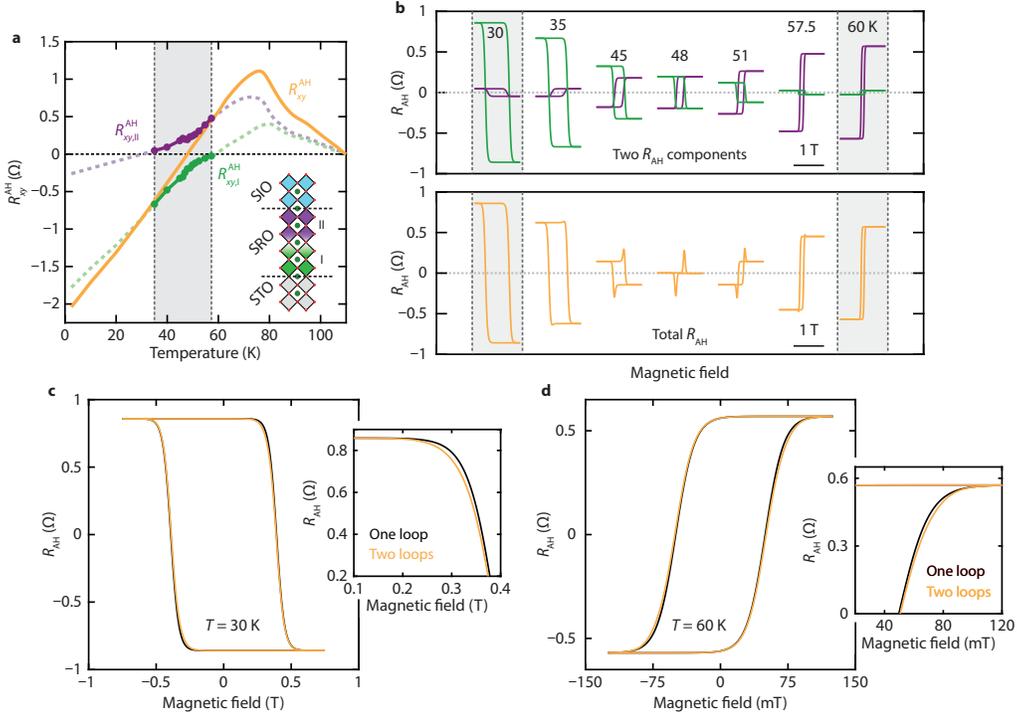

FIG. S7: **Extrapolation of the AH channels. a,** Total $R_{xy}^{\mathrm{AH}}$ and the extracted $R_{xy}^{\mathrm{AH}}$ from the two anomalous Hall components. The dashed lines illustrate a possible temperature dependence of $R_{xy}^{\mathrm{AH}}$. **b,** Two anomalous Hall components (top) that add up to the total $R_{\mathrm{AH}}$ curves (bottom). At 35 and 60 K, the two components cannot be discerned and the smaller component is based on extrapolation of the data in panel **a. c, d,** Comparison between a single AHE and a total AHE resulting from two anomalous Hall components of the same sign at 30 and 60 K, respectively.

### D. Partial switching of the spin polarization

In Fig. S8 we present additional measurements on the partial switching of the spin polarization of the two channels. Figure S8a shows simulated and measured $R_{AH}$ curves similar to those in Fig. 4c. However, in this case the measurement starts from the bottom branch (opposite spin polarization) and is traversed in the opposite direction (in the range $[0, -B_{max}, B_{max}, 0]$). As expected, the measurements are mirrored with respect to those with opposite spin polarization. In Fig. S8c we show that the offset of the center of the hysteresis loop is in the opposite direction. The shift of the loop is about $37\,\mathrm{mT}$ at $45\,\mathrm{K}$ (see Fig. S8b). At higher $B_{max}$, the shift of the center of the loop becomes smaller, which can be understood by the partial switching of polarization of the layer with larger $B_c$ (see Fig. S8d).

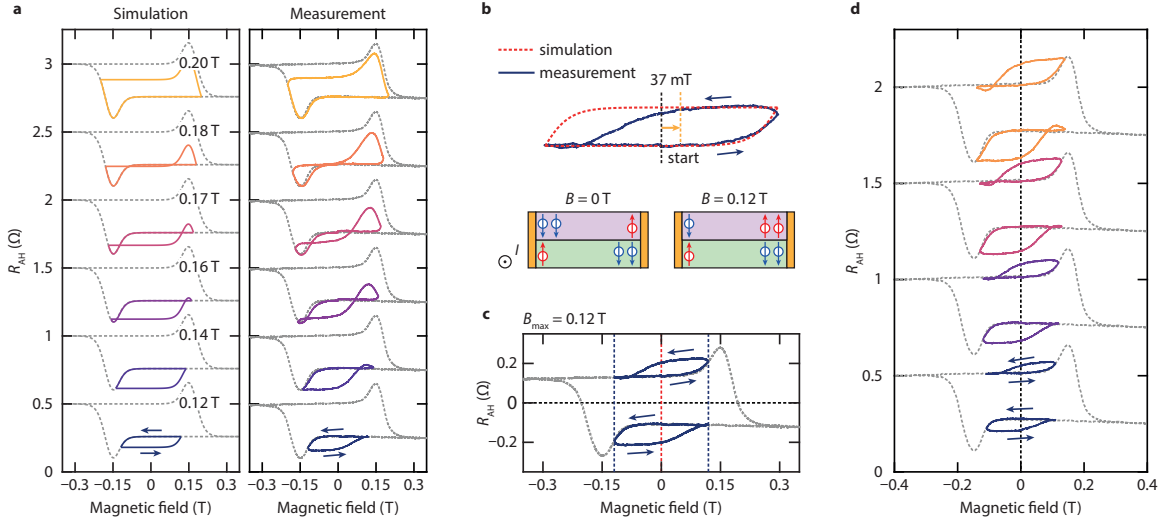

FIG. S8: **Control of the spin polarization of the two channels. a,** Simulated (left) and measured (right) $R_{AH}$ curves. The curves are offset vertically. **b,** Top: comparison of simulated and measured $R_{AH}$ for $B_{max} = 0.12\,\mathrm{T}$. Bottom: schematic of the microscopic picture. **c,** Measured $R_{AH}$ starting with positive or negative spin polarization. **d,** Measured $R_{AH}$ starting with positive or negative spin polarization. The curves are offset vertically.

### E. Variation of the SIO thickness

In Fig. S9a, we show the sheet resistance versus temperature for an STO/SRO/SRO heterostructure, as well as a number of heterostructures with different SIO layer thicknesses. We find that increasing the number of SIO layers lowers the resistance, indicating that the SIO layer acts as a parallel resistor. However, the sign reversal of the anomalous Hall effect (Fig. S9b) does not appear to depend on the number of SIO layers, indicating that the properties of the SRO are affected by the interfacial SIO layer. In addition, the residual resistance ratio (RRR = $R_{300\,\mathrm{K}}/R_{1.5\,\mathrm{K}}$) is the same for all STO/SRO/SIO heterostructures and lowered with respect to the STO/SRO/STO.

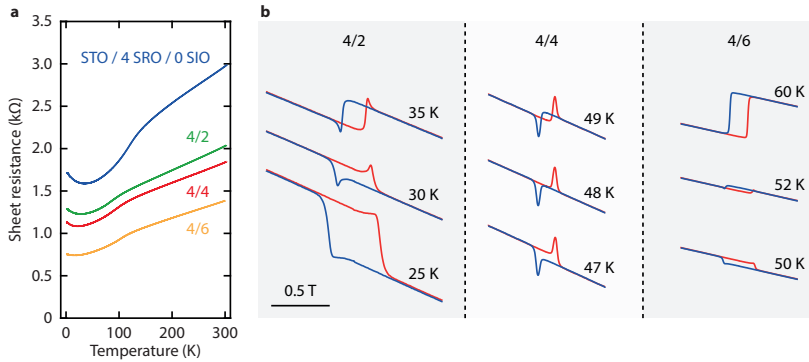

FIG. S9: **Variation of the SIO layer thickness. a,** Sheet resistance and **b,** $R_{xy}$ of STO/SRO/SIO heterostructures with different SIO thicknesses versus temperature.

## III. DENSITY FUNCTIONAL THEORY CALCULATIONS

### A. Computational details

Atomic relaxation was performed for all the considered systems. The internal degrees of freedom were optimized by minimizing the total energy to be less than $10^{-5}$ eV and the remaining forces to be less than 10 meV Å$^{-1}$. A $6 \times 6 \times 1$ $k$-point centered in $\Gamma$ was used for the calculation of the heterostructure, while a $10 \times 10 \times 8$ was used for the phases without rotations. We used a $8 \times 8 \times 1$ k-point grid for the calculation of the density of states (DOS). For the estimate of the $D/J$ ratio, a $2a \times 2a \times 2c$ supercell and an $8 \times 8 \times 8$ $k$-point grid were used.

The Hubbard $U$ effects on the Ru and Ir sites were included within the LSDA+U approach using the rotational invariant scheme [4]. In the study of the STO/SRO/SIO heterostructure, a small value of $U = 0.2$ eV on the Ru site was used to understand the qualitative behaviour of the magnetic profile, while a larger value of $U = 1.4$ eV for the Ru and $U = 1.2$ eV for the Ir was used in the non-collinear calculations to get numerical stability in the calculation of the Dzyaloshinskii-Moriya interactions. The Hund's coupling $J_H$ was always set equal to $0.15U$.

### B. DFT study of the STO/SRO/SIO heterostructure in a collinear approximation

DFT calculations were performed for the STO/SRO/SIO heterostructure. There are two different interfacial layers within the SRO: the SRO layer interfaced with STO and the SRO layer interfaced with SIO. We will show that the inner layers, the STO interfacial layer and the SIO interfacial layer have different properties; however, no major qualitative changes are observed in the heterostructure with respect to the bulk since the magnetic order, the structural phase and the DOS are barely affected.

From the structural point of view, SRO and SIO belong to the same space group. However, epitaxial constraint can induce different octahedral tilts and distortions with respect to the bulk. In this case, this results in a complete suppression of octahedral rotations as shown in Fig. S3b.

When a large value of the Coulomb repulsion is used, the magnetic moment saturates and tends to be similar throughout the heterostructure. When an intermediate value of the Coulomb repulsion is used, the magnetic profile (Fig. S10, right panel) shows that the interfacial layers present a magnetic proximity effect. This results in a reduction of the Ru magnetic moment of the interfacial layers as observed in other ruthenate systems. The magnetic moments of the STO and SIO interfacial layers are similar. However, in the real heterostructure, the diffusion of the Ir

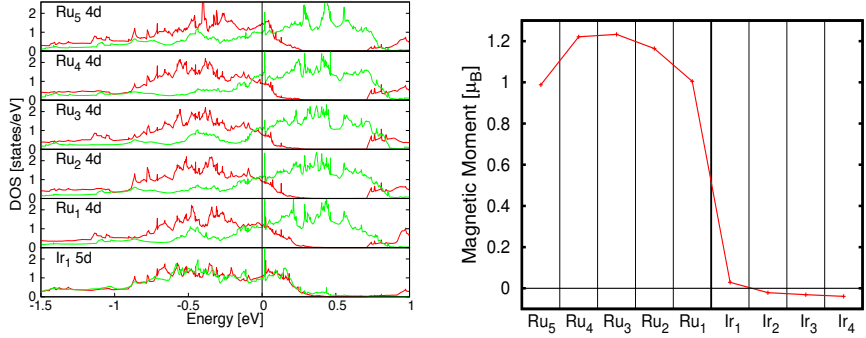

FIG. S10: **Layer-projected DOS and magnetization.** Left: layer-projected DOS for the *d*-states of the Ru layers and the Ir interfacial layer as a function of the distance from the interface. Right: layer-dependent magnetization of an STO(3)/SRO(5)/SIO(4) heterostructure.

atoms in the Ru layers can further reduce the Curie temperature and the magnetic moment of the SIO interfacial layers.

The two interfacial layers display different peculiarities from the electronic point of view. Our calculations show that the interfacial STO layer presents a bandwidth reduction due to the decrease in connectivity [5, 6] while the interfacial SIO layer induces large spin–orbit coupling. The layer projected DOS does not show sufficiently strong modifications at the Fermi level to induce a weakening of the Stoner ferromagnetism, which happens when the Ru states are localized [7]. Despite the absence of qualitative changes, many quantitative changes are observed, and the anomalous Hall conductivity strongly depends on the way in which these changes influence the band structure.

### C. Non-collinear study of the STO/SRO/SIO heterostructures

To determine the magnitude of the $D/J$ ratio at the interface, we performed non-collinear calculations. We first focus on the strength of the magnetism shown in Fig. S11. First-principles calculations show that the $SRO_n/SIO_{6-n}$ system is metallic and ferromagnetic for all values of $n$. The ferromagnetism is weakened in the ultrathin limit, but for 4-5 monolayers ($n = 4, 5$) the energy difference of the superlattice tends to the value for bulk SRO. The energy difference per Ru atom depends on the connectivity and the magnetic exchange $J$, while the energy difference per Ru-Ru bond is proportional to the average of $J$. The energy difference per Ru atom at $n = 1$ is almost half of the bulk value, while the energy difference per Ru-Ru bond is circa 20% smaller than the bulk value.

Using the symmetry analysis reported in literature for perovskites [8], we observe that there are two inequivalent DM vectors in bulk SRO: $D_c$ is the DM vector between two Ru atoms on different planes, while $D_{ab}$ is the DM vector between two Ru atoms lying on the same plane. From our non-collinear calculations for the SRO bulk we obtain: $\vec{D}_{ab}^{\mathrm{Bulk}} = (2.17, 1.17, 2.37)$ meV and $\vec{D}_c^{\mathrm{Bulk}} = (0.62, 3.34, 0)$ meV. Calculating the ratio between the modulus of the $D$ vectors and the exchange $J$, we obtain $D_{ab}^{\mathrm{Bulk}}/J_{ab}^{\mathrm{Bulk}} = 0.186$ and $D_c^{\mathrm{Bulk}}/J_c^{\mathrm{Bulk}} = 0.163$. Hence, the Ru atoms have a moderately large intrinsic $D/J$ ratio.

We study the case of the SRO(1)/SIO(1) superlattice without relaxation in order to investigate the bare effect of the Ir atoms on the $D/J$ ratio. For the DM vector we obtain $\vec{D}_{ab}^{\mathrm{SRO/SIO}} = (1.98, 1.26, 2.75)$ meV and $D_{ab}^{\mathrm{SRO/SIO}}/J_{ab}^{\mathrm{SRO/SIO}} = 0.236$. It was suggested that a $D/J$ ratio in the *ab* plane larger than 0.3 can explain the experimental results at the SRO/SIO interface in terms of skyrmions [9]. Since the $D/J$ ratio does not reach 0.3 for this system, the increase will be not large enough to determine a magnetic phase change from ferromagnetism to skyrmions.

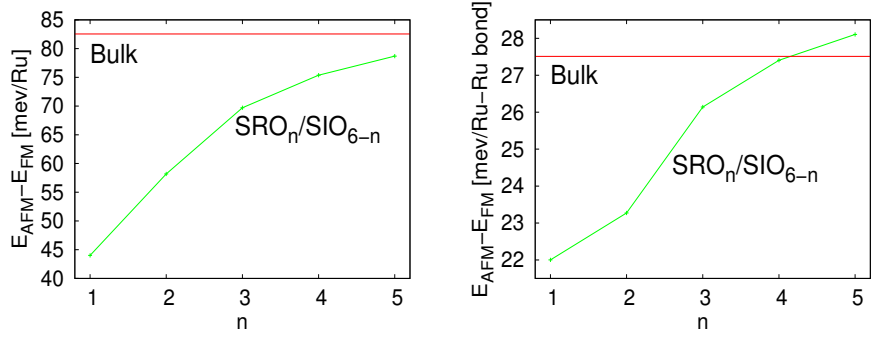

FIG. S11: **Magnetism in the SRO$_n$/SIO$_{6-n}$ system.** Evolution of the energy difference between the ferromagnetic and the G-type antiferromagnetic phase as function of the number of SRO layers $n$ for the superlattice SRO$_n$/SIO$_{6-n}$ in the non-collinear calculation. The bulk value is shown as red, while the value of the heterostructure as green. In the left panel we report the energy difference per Ru atom, while in the right panel we report the energy difference per Ru-Ru bond.

## IV. ELECTRONIC PROPERTIES OF THE SRO/STO AND SRO/SIO INTERFACES

We study the electronic properties of the SRO/SIO and SRO/STO interfaces in the tetragonal phase without octahedral rotations. Using the Wannier function method [10], we determine the effective hopping parameters for the $t_{2g}$ manifold of the transition metal atoms.

### A. Tight-binding model for the superlattice without octahedral distortions

A tight-binding model is derived including the nearest-neighbor (NN) and next-nearest-neighbor (NNN) hopping terms without the inclusion of SOC; the latter, however, will be included afterwards in the Hall conductivity calculation. We restrict our analysis to the effective $d$-bands of the transition metal (M) atoms. We simulate a bulk system with 2 metal atoms in the unit cell, but we write the tight-binding model for a bilayer without performing the Fourier transform along $k_z$. We label the two atoms of the bilayer as 1 and 2. The Hamiltonian in matrix form of the bilayer is

$$\hat{H}(k_x, k_y) = \begin{pmatrix} H_{11}(k_x, k_y) & H_{12}(k_x, k_y) \\ H_{21}(k_x, k_y) & H_{22}(k_x, k_y) \end{pmatrix}.$$

Each submatrix is a sum of the different hopping related to the energy on site, NN and NNN as $H_{11} = H_{11}^0 + H_{11}^{NN} + H_{11}^{NNN}$, where

$$H_{11}^0 = \begin{pmatrix} \varepsilon_{xy}^0 & 0 & 0 \\ 0 & \varepsilon_{xz}^0 & 0 \\ 0 & 0 & \varepsilon_{yz}^0 \end{pmatrix},$$

$$H_{11}^{NN} = \begin{pmatrix} 2t_{xy,xy}^{100}\cos k_x a + 2t_{xy,xy}^{010}\cos k_y a & 0 & 0 \\ 0 & 2t_{xz,xz}^{100}\cos k_x a + 2t_{xz,xz}^{010}\cos k_y a & 0 \\ 0 & 0 & 2t_{yz,yz}^{100}\cos k_x a + 2t_{yz,yz}^{010}\cos k_y a \end{pmatrix},$$

$$H_{11}^{NNN} = \begin{pmatrix} 4t_{xy,xy}^{110}\cos k_x a \cos k_y a & 0 & 0 \\ 0 & 4t_{xz,xz}^{110}\cos k_x a \cos k_y a & -4t_{xz,yz}^{110}\sin k_x a \sin k_y a \\ 0 & -4t_{yz,yz}^{110}\sin k_x a \sin k_y a & 4t_{yz,yz}^{110}\cos k_x a \cos k_y a \end{pmatrix},$$

while for the submatrix $H_{12}$ connecting the two layers we have

$$H_{12}^{NN} = \begin{pmatrix} t_{xy,xy}^{001} & 0 & 0 \\ 0 & t_{xz,xz}^{001} & 0 \\ 0 & 0 & t_{yz,yz}^{001} \end{pmatrix},$$

$$H_{12}^{NNN} = \begin{pmatrix} 2t_{xy,xy}^{101}\cos k_x a + 2t_{xy,xy}^{011}\cos k_y a & 2it_{xy,xz}^{011}\sin k_y a & 2it_{xy,yz}^{101}\sin k_x a \\ 2it_{xz,xy}^{011}\sin k_y a & 2t_{xz,xz}^{101}\cos k_x a + 2t_{xz,xz}^{011}\cos k_y a & 0 \\ 2it_{yz,xy}^{101}\sin k_x a & 0 & 2t_{yz,yz}^{101}\cos k_x a + 2t_{yz,yz}^{011}\cos k_y a \end{pmatrix}.$$

The breaking of inversion symmetry along $z$ (BISz) gives rise to spin-independent hopping terms that mix $xy$ and $\gamma z$ orbitals also for the NN. These new types of hopping are responsible for the Rashba effect [11]. For the purposes of our present analysis such terms are negligible, being of the order of few meV and modifying the band structure far (0.5 eV) above the Fermi level.

$$H_{11}^{NN,\text{BISz}} = \begin{pmatrix} 0 & 2it_{xy,xz}^{010}\sin k_y a & 2it_{xy,yz}^{100}\sin k_x a \\ -2it_{xy,xz}^{010}\sin k_y a & 0 & 0 \\ -2it_{xy,yz}^{100}\sin k_x a & 0 & 0 \end{pmatrix},$$

$$H_{11}^{NNN,\text{BISz}} = \begin{pmatrix} 0 & 4it_{xy,xz}^{110}\sin k_y a\cos k_x a & 4it_{xy,yz}^{110}\sin k_x a\cos k_y a \\ -4it_{xy,xz}^{110}\sin k_y a\cos k_x a & 0 & 0 \\ -4it_{xy,yz}^{110}\sin k_x a\cos k_y a & 0 & 0 \end{pmatrix}.$$

**B. Hopping parameters for the SRO/SIO and SRO/STO superlattices**

The band structure of the tetragonal SRO/SIO system is presented in Fig. S12. The total bandwidth is 3.8 eV and the states around the Fermi level are predominantly Ru and Ir $t_{2g}$ states.

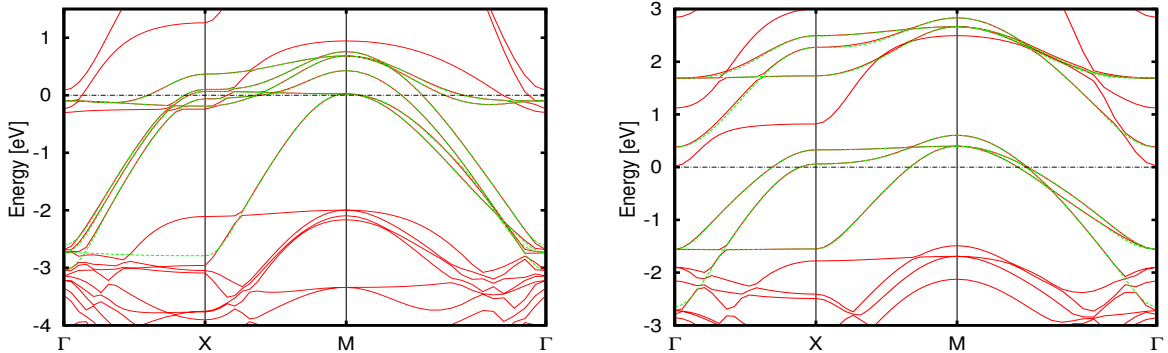

FIG. S12: **Band structure of SRO/SIO and SRO/STO.** DFT band structure (red) and $t_{2g}$ bands (green) obtained using the Wannier functions for the nonmagnetic phase of SRO/SIO (left panel) and SRO/STO (right panel) superlattices. The Fermi level is set to zero.

The band structure agrees with the reported band structure of tetragonal ruthenates and the hopping parameters are quantitatively similar to those previously reported for the ruthenates [12].

We list the hopping parameters between Wannier functions (WF) of the same species in Table I and between the two different species in Table II. In the tetragonal case there is no first neighbour hopping between different orbitals. Non-diagonal hoppings can be activated by the octahedral rotations. From the on-site energy, we can see that the crystal field splitting of the Ru has opposite sign with respect to the Ir, while the on-site energy is similar for all the considered orbitals.

|  | on site | NN | | NNN |
|---|---|---|---|---|
|  | 000 | 100 | 010 | 110 |
| xy(Ir)-xy(Ir) | 0 | -381 | -381 | -139 |
| xz(Ir)-xz(Ir) | -15 | -374 | -27 | 18 |
| yz(Ir)-yz(Ir) | -15 | -27 | -374 | 18 |
| xz(Ir)-yz(Ir) | 0 | 0 | 0 | 21 |
| xy(Ru)-xy(Ru) | 223 | -346 | -346 | -125 |
| xz(Ru)-xz(Ru) | 182 | -374 | -26 | 13 |
| yz(Ru)-yz(Ru) | 182 | -26 | -374 | 13 |
| xz(Ru)-yz(Ru) | 0 | 0 | 0 | 17 |

TABLE I: Hopping integrals between the M atoms for the nearest-neighbor (NN), for the next-nearest-neighbor (NNN) as the selected WFs of SRO/SIO between WFs of the same atomic specie. The on-site energy of the SIO xy-like WF is set to zero. The unit is meV.

The Ru and Ir hoppings are quite similar and one can separate the hopping parameters in 4 groups. The NN $\pi$-hoppings are between $-381$ and $-318$ meV. The NN $\delta$-hoppings are between $-27$ and $-25$ meV. For the NNN case, we find a mixing of $\sigma$, $\pi$ and $\delta$ bonds. The NNN hopping parameters with mainly $\sigma$-component are between $-139$ and $-125$ meV. Other NNN hopping are between 13 and 21 meV. Octahedral rotations would strongly decrease the hopping parameters, reducing the bandwidth [12] and increasing the magnetic properties, but would be qualitatively different. We note that the second neighbour hopping between $xz(\text{Ir})$-$yz(\text{Ir})$ is 21 meV along the (110) direction, while the hopping between $xz(\text{Ru})$ and $yz(\text{Ru})$ is 17 meV. We will show below that these terms are crucial for the topological properties of the system.

The two main differences of the $t_{2g}$ Ti states with respect to the Ru states are their position of circa 3 eV above the $t_{2g}$ Ru states and a smaller size of the hopping due to the reduced atomic radius of the Ti. At odds with the previous case, the Ru and Ti bands are well separated in energy as shown in Fig. S12, while in the SRO/SIO superlattice there is a strong mixing between the $d$-states of the different atoms. We list the hopping parameters between the same species in Table III and between the two different species in Table IV. We observe that the $xz(\text{Ti})$-$yz(\text{Ti})$ along (110) is just 5 meV, much smaller than the $xz(\text{Ru})$-$yz(\text{Ru})$ and $xz(\text{Ir})$-$yz(\text{Ir})$ hopping.

|  | NN | NNN | |
|---|---|---|---|
|  | 001 | 101 | 011 |
| xy(Ru)-xy(Ir) | -25 | 14 | 14 |
| xz(Ru)-xz(Ir) | -318 | -125 | 14 |
| yz(Ru)-yz(Ir) | -318 | 14 | -125 |
| xy(Ru)-xz(Ir) | 0 | 0 | 16 |
| xy(Ir)-xz(Ru) | 0 | 0 | 18 |

TABLE II: Hopping integrals between the M atoms for the nearest-neighbor (NN), for the next-nearest-neighbor (NNN) as the selected WFs of SRO/SIO between WFs of different atomic species. The unit is meV.

| | on site | NN | | NNN |
|---|---|---|---|---|
| | 000 | 100 | 010 | 110 |
| xy(Ti)-xy(Ti) | 2344 | -286 | -286 | -83 |
| xz(Ti)-xz(Ti) | 2321 | -301 | -40 | 2 |
| yz(Ti)-yz(Ti) | 2321 | -40 | -301 | 2 |
| xz(Ti)-yz(Ti) | 0 | 0 | 0 | 5 |
| xy(Ru)-xy(Ru) | 0 | -348 | -348 | -126 |
| xz(Ru)-xz(Ru) | 121 | -306 | -22 | 15 |
| yz(Ru)-yz(Ru) | 121 | -22 | -306 | 15 |
| xz(Ru)-yz(Ru) | 0 | 0 | 0 | 19 |

TABLE III: Hopping integrals between the M atoms for the nearest-neighbor (NN), for the next-nearest-neighbor (NNN) as the selected WFs of SRO/STO between WFs of the same atomic specie. The on-site energy of the SRO xy-like WF is set to zero. The unit is meV.

| | NN | NNN | |
|---|---|---|---|
| | 001 | 101 | 011 |
| xy(Ru)-xy(Ti) | -30 | 11 | 11 |
| xz(Ru)-xz(Ti) | -314 | -112 | 11 |
| yz(Ru)-yz(Ti) | -314 | 11 | -112 |
| xy(Ru)-xz(Ti) | 0 | 0 | 16 |
| xy(Ti)-xz(Ru) | 0 | 0 | 7 |

TABLE IV: Hopping integrals between the M atoms for the nearest-neighbor (NN), for the next-nearest-neighbor (NNN) as the selected WFs of SRO/STO between WFs of different atomic species. The unit is meV.

## V. CALCULATION OF THE BERRY CURVATURE

### A. Computational details for the Berry curvature calculation of the bilayers

To estimate the Berry curvature, we determine the amplitude of the hopping parameters of the $t_{2g}$ electrons in the tetragonal phase without octahedral distortions for the SRO/SIO and SRO/STO superlattices. From the electronic point of view, the SRO at the STO interface shows an increased electronic localization compared to the inner layers, while the SIO interface layer shows a large SOC. The low-energy electronic structure is constructed on the $t_{2g}$ bands of the Ru, Ir, and Ti atoms and the orbital-dependent connectivity matrix between first and second nearest neighbors is taken from the ab-initio calculations. In order to single-out the electronic terms which can be relevant for getting a non-trivial Berry curvature we consider a model system composed of Ru/Ru, Ru/Ir and Ru/Ti bilayers. The bilayer model is able to capture the differences between the electronic behavior of the SRO/SIO interface, the SRO/STO as well as in the inner layers of the ferromagnetic SRO. Such basic configurations can effectively mimic the electronic behavior of the inner layers of the SRO, and of the SRO/SIO and SRO/STO interfaces. The ferromagnetic state is simulated by adding an intra-orbital interaction that can lead to a net magnetization along the $z$-axis. We can capture the main features of the Berry phase while neglecting the octahedral distortions, and the results are not sensitive to a change of the easy axis of the magnetization.

The intrinsic contribution to the anomalous Hall conductivity is dependent only on the topological properties of band structure of the crystal through the Berry curvature of the Bloch bands as:

$$\sigma_{ij}^{AH-int} = -\varepsilon_{ijl}\frac{e^2}{\hbar}\int\frac{1}{(2\pi)^d}d\mathbf{k}f(E_n(\mathbf{k}))\Omega_n(\mathbf{k}).$$

where $d$ is the system dimensionality, $\varepsilon_{ijl}$ is the Levi-Civita tensor, $a_n(\mathbf{k})$ is the Berry connection $a_n(\mathbf{k}) = i\langle n_{\mathbf{k}}|\nabla_{\mathbf{k}}|n_{\mathbf{k}}\rangle$, and $\Omega_n(\mathbf{k})$ is the Berry-phase curvature, $\Omega_n(\mathbf{k}) = \nabla \times a_n(\mathbf{k})$, corresponding to the eigenstate $|n_{\mathbf{k}}\rangle$. The Berry curvature is then computed from the eigenstates $|n_{\mathbf{k}}\rangle$ for each band assuming a sufficiently fine mesh in the Brillouin zone [13]. The agreement with experiments is good at low magnetization while at intermediate/high magnetization the effects of the octahedral rotations of bulk SRO should be taken in account to have an agreement [14].

## B. Symmetry analysis and Berry curvature for the SRO single layer

The results can be understood by considering the symmetry properties of the $t_{2g}$ electronic structure for a SRO single layer in more detail (see Fig. S13). Apart from the spatial symmetries, the examined quantum problem has an important internal symmetry, i.e., spin-orbital parity. The spin and orbital terms in the Hamiltonian are such that the parity of the sum of the quantum numbers associated with the projections of the orbital $L = 1$ and spin $S = 1/2$ angular momenta is conserved. Without an external field or intrinsic magnetization, the parity is conserved for any direction in the spin-orbital space, thus leading to a Kramers degeneracy. In the presence of a source of time reversal symmetry breaking, one can still employ the spin-orbital parity with the constraint of selecting only the direction of the external field or intrinsic magnetization. Hence, starting from the six local states (i.e., 3 orbitals and 2 spin configurations for each transition metal element) and employing the spin-orbital parity symmetry, one can separate the Hamiltonian at any given $\mathbf{k}$ of the Brillouin zone into $3 \times 3$ blocks. The basis of the positive-parity sector is spanned by the 3 spin-orbital states $|\sigma^z, L^z\rangle$; $|\downarrow, +1\rangle$, $|\downarrow, -1\rangle$ and $|\uparrow, 0\rangle$. Within this representation, the electronic structure can be investigated separately in each sector.

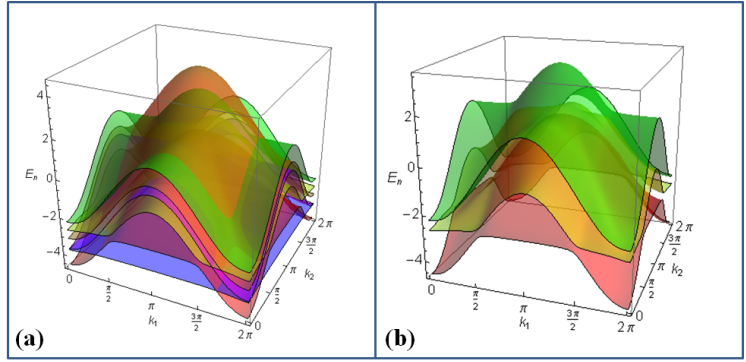

FIG. S13: **Single layer electronic structure close to the Fermi level for the Ru $t_{2g}$ bands. a**, There are six $t_{2g}$ bands close to the Fermi level. **b**, Projected bands with a given quantum number for the spin-orbital parity.

The key finding here is that the Berry connection for the bands in each parity sector is non-trivial and its integrated value yields a non vanishing Chern number $\mathcal{C}$. In particular, from an inspection of the three bands forming each parity block, two bands with $d_{xz}, d_{yz}$ character have $\mathcal{C}=\pm2$ and another one with mainly $d_{xy}$ nature having $\mathcal{C} = 0$. The sum of the Chern numbers in each parity sector is zero as expected by symmetry arguments. Furthermore, the Berry curvature within the other spin-orbital parity sector has opposite signs and the same distribution of Chern numbers

$\pm 2, 0$. Thus, in the absence of an external field or intrinsic magnetization the bands with opposite Chern numbers fall on top of each other and become trivial. A representative case of topologically non-trivial bands is reported in Figs. S14(a-c) showing Berry curvature profiles $\Omega_n(\mathbf{k})$ for $n = 1, 2, 3$ within positive parity sector. Plots (a) and (b) with sharp peaks correspond to bands with non-vanishing Chern numbers whereas plot (c) refers to the trivial band, which can however still contribute to the total anomalous Hall conductivity. The inspection of the Hamiltonian structure indicates that the inter-orbital next nearest-neighbor hopping $t_{xz,yz}^{110}$ is a crucial parameter to drive a topological transition with a Berry curvature having non-zero Chern number for the tangled bands in each parity sector. Its role is both to open the indirect gap between the topologically non-trivial bands, roughly at the positions of the peaks in Figs. S14(a-b), and to break remaining orbital time-reversal symmetry in the fixed parity sectors. It was shown that this kind of term can produce AHC in $e_g$ partially-filled transition metal perovskites [15].

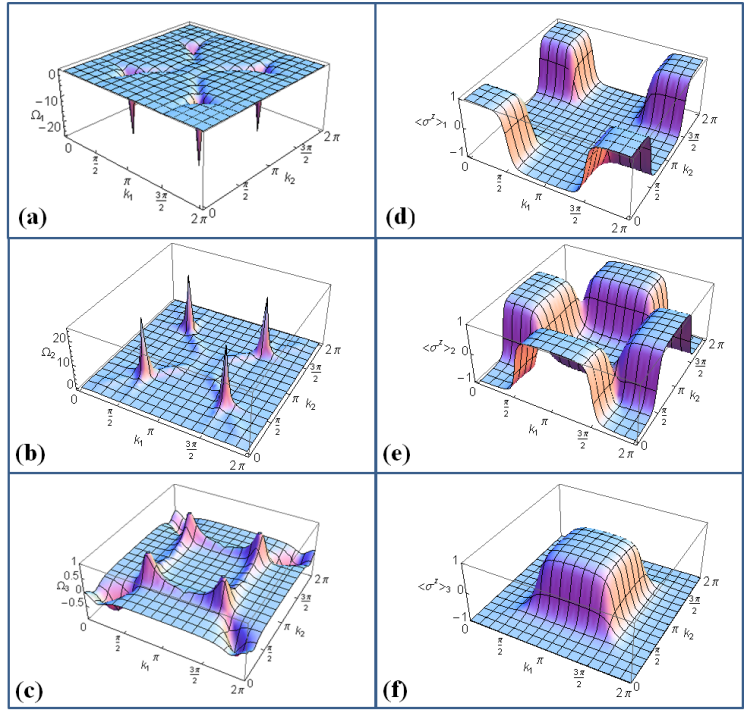

FIG. S14: **Berry curvatures.** Left column, **a–c**, Berry curvatures $\Omega_n(\mathbf{k})$ associated with the bands $n = 1, 2, 3$ reported in Fig. S13b. The two lowest energy bands (a-b) are topological non-trivial with a Berry curvature that results in Chern numbers $\mathcal{C} = \pm 2$. The third band has a non-vanishing amplitude of the Berry curvature within the Brillouin zone (c), however the sign changes gives $\mathcal{C} = 0$. Right column, **d–f**, Spin polarizations $\langle \sigma^z \rangle_n$ for the corresponding bands. Note that in each case the variation of $\langle \sigma^z \rangle_n$ is the full range of $[-1, 1]$.

The topological non-triviality of the bands in the parity sectors is related with a sign change of their spin polarizations $\langle \sigma^z \rangle_n \equiv \langle n_\mathbf{k} | \sigma^z | n_\mathbf{k} \rangle$. Figs. S14(d-f) show the profiles of $\langle \sigma^z \rangle_n$ for $n = 1, 2, 3$ bands within the topological regime. It is notable that these polarizations interpolate between two extreme values of $\langle \sigma^z \rangle = \pm 1$ as one moves across the Brillouin zone. This effect does not occur in the trivial regime, which implies that the band inversion leading to topological non-triviality is tied to the sign change of the spin-polarization and, since the orbital state is locked with the spin in the fixed parity sector, the sign change also occurs in the orbital space. Hence, the observed topological states are triggered by the spin-orbital fluctuations and the main control parameter driving the topological phase transition is the value of the external field or intrinsic magnetization. Indeed, one finds that if its magnitude is more than twice as large as the in-plane hopping amplitudes, then the spin mixing shown in Figs. S14(d-f) vanishes and the bands become trivial.

Moreover, it is the band spin-polarization that leads to a splitting of the parity sectors such that no cancellation between the opposite parity blocks occur. Then, since we are dealing with a ferromagnetic metallic system and each parity sector is partially filled, we point out that a complete cancellation for the Berry curvature associated to the occupied bands can be only accidental. The occupation of the two active bands in each parity sector and the related effects from the spin-splitting between them, due to the magnetization, are the key mechanisms for decoding the different orbital and spin channels in the anomalous Hall conductivity. Due to the intrinsic topological behavior of the spin-orbital entangled bands, the contribution of the Berry curvature to the conductivity is weakly dependent on the details of the electronic structure as far as there are no band crossings in each parity sector or symmetry breaking effects.

The proximity to an inequivalent layer with non-magnetic Ti or Ir t$_{2g}$ bands close to the Fermi level does not alter the topological behavior of the Ru bands but rather influences their relative occupation number. In particular, since Ti is in a $d^0$ configuration, the hybridization of the Ru/Ti bands is negligible as well as the magnetic proximity. This implies that the Ti layer is not active until a value for the Ru magnetization is reached that allows the minority spin electrons to leak into the Ti bands. In that regime, one can drive again a sign change in the conductivity. The Ru/Ru and Ru/Ir systems are more complex due to the interference of topological non-trivial bands which can fully contribute to the Berry curvature. This is always the case for the Ru/Ru bilayer, while it is activated by the magnetic proximity for the Ru/Ir bilayer. The different weights of the topologically active t$_{2g}$ bands can account for the inequivalent behavior of the conductivity.

## C. Anomalous Hall conductivity of the bilayer systems

The anomalous Hall conductivity for the Ru/Ir, Ru/Ru, and Ru/Ti bilayers, presented in Fig. 1h, is remarkably different in the low-to-intermediate magnetization regime, being positive for Ru/Ir, negative for the Ru/Ti and changing sign (positive-to-negative) for the Ru/Ru bilayer. This result provides the first indications of the intrinsic competition which can manifest in the AH conductivity when comparing the STO-SRO-STO and SIO-SRO-SIO heterostructures. One can immediately observe that the contribution of the SIO to the AH conductivity, due to the induced magnetic moment of the SRO into the SIO, is positive for any amplitude of the Ru magnetic moment. The Ir/Ru mixing of the electronic states at the interface is thus able to completely counterbalance the negative AH conductivity arising uniquely from the two-dimensional Ru bands since they are decoupled from the Ti states for the Ru/Ti interface. Here, it is worth pointing out that, while the electronic structure of the SIO at the interface has the same topological character as that in the SRO, there are significant differences in the spectra emerging from the (i) amplitude of the magnetization in the SIO layers, (ii) the filling of the t$_{2g}$ bands, and (iii) the larger spin-orbit coupling of Ir compared to Ru. At this stage, one would then qualitatively conclude that the Ru bands at the interface are more prone to provide an overall negative AH contribution while the presence of the Ir states can turn the AH conductivity from negative to positive. Here, the regime of small magnetization is physically relevant at the interface where the ferromagnetic moment is expected to be reduced with respect to the inner layers as also demonstrated by the ab-initio simulations. On the other hand, the Ru/Ru bilayer exhibits a positive-to-negative sign change when moving from a small to intermediate magnetization regime.

### D. Berry curvature of STO/SRO/STO and STO/SRO/SIO heterostructures

To study the Berry curvature of our multilayer system, we consider heterostructures of the type STO($k$)/ SRO($n$)/STO($m$) and STO($k$)/SRO($n$)/SIO($m$), where the number of layers is $k = 2$, $m = 2$ and $n = 4$ (see Fig. S15a). The analysis is based on an effective model Hamiltonian that captures the electronic behavior close to the Fermi level in the inner layers of the SRO and at the STO and SIO interfaces. The electronic parameters for the $t_{2g}$ bands are provided by the DFT analysis as implemented in the bilayer calculation. To assess the role of an inhomogeneous magnetization profile, we simulated different layer-dependent magnetic configurations. Such a profile is consistent with the expectation of a different magnetocrystalline anisotropy at the STO/SRO and SRO/SIO interfaces due to the inequivalent tetragonal distortions and variation of the spin–orbit coupling strength. Taking into account the input from the DFT, we assume that the magnetization of the SRO is largest in the inner layers and decreases at the STO and SIO interfaces. Due to the weak tetragonal distortions, we assume that the magnetization is orbital-independent. Since the $c$-axis is the easy-axis direction for the SRO magnetization, we take this configuration as the most favorable and consider distinct layer dependences of the magnetization amplitude for the whole heterostructure. To simulate the temperature dependence of the AHE in the ferromagnetic phase, the thermal magnetization in each layer is described by the Stoner model for itinerant ferromagnets as $M(T) = M_0[1 - (\frac{T}{T_C})^2]^{1/2}$, where $M_0$ is the magnetization at zero temperature and $T_C$ the Curie temperature.

While there is no significant magnetic reconstruction at the STO/SRO interface, the SRO/SIO interface can exhibit various types of magnetic reconstructions with a small induced Ir magnetic moment. In order to evaluate how this influences the Berry curvature and the anomalous Hall effect, we have explored the possibility of (i) nonmagnetic SIO, (ii) a collinear ferromagnetic proximity effect in the interfacial SIO layers, (iii) a reorientation of the magnetic moments in the interfacial SIO layers with a magnetization component lying in the $ab$ plane, and (iv) the combination of (ii) and (iii) with an effective canting of the magnetization in the interfacial SRO layers due to the combination of $c$-axis and $a$-axis spin polarizations. We note that increasing the thickness of the SRO subsystem restores the sign change of the AH conductivity because the weight of the interfacial layer is reduced with respect to that of the inner SRO layers.

The main outcomes are reported in Figs. S15 and S16. We first consider the STO/SRO/STO heterostructure. Figure S15b shows that the AHE has an intrinsic tendency to change sign as a function of temperature. We also find that changing the amplitude of the magnetization leads to a shift of the effective temperature below which the AHE changes sign. This indicates that there are different electronic channels with opposite Berry curvatures, with weights dependent on the strength of the magnetization. Small magnetizations are associated with a positive AHE, while a large amplitude of the Ru moments can drive a sign change of the AH conductivity.

In the STO/SRO/SIO system [Fig. S15(c)], the contribution from the SIO/SRO interface tends to oppose to the sign change of the SRO system. The asymmetric system displays a sign change for $M_0 \geq 1.4$ $\mu_B$ (green and blue curves). For smaller $M_0$, the SIO contribution tends to keep $\sigma_{xy}$ positive in the entire temperature range (red curve). This trend is consistent with the bilayer description.

We now investigate the evolution of $\sigma_{xy}$ in the presence of a induced magnetization on the Ir site, as shown in Fig. S16(a). We first consider that a small, collinear Ir magnetic moment is induced in the SIO layers close to the interface due to the ferromagnetic proximity effect. In this case, the AHE displays a clear sign change for $M_0 \geq 1.4$ $\mu_B$. Otherwise, the SIO contribution leads to a dominant positive sign of the conductivity (red curve).

Finally, we investigate the possibility of having an in-plane component due to interface-driven

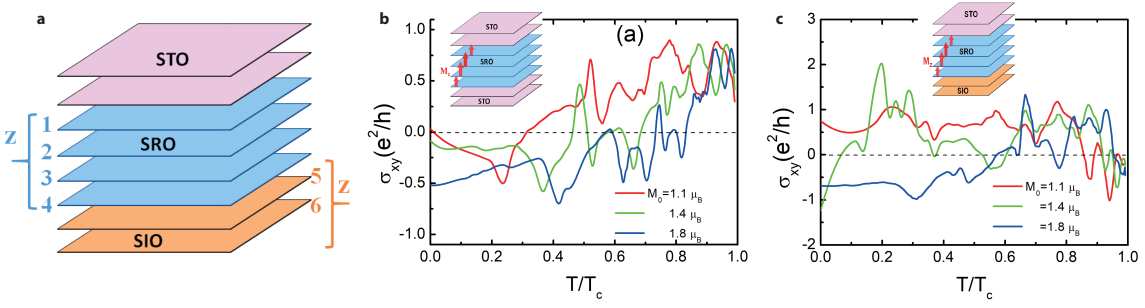

FIG. S15: **Temperature dependence of the AH conductivity. a**, Schematic of the simulated heterostructure, where $z$ represents the layer number. **b-c**, $\sigma_{xy}$ versus $T/T_c$ for STO/SRO/STO (b) and SIO/SRO/STO (c) heterostructures. We consider the magnetization to be parallel to the $c$-axis with an inhomogeneous layer profile within the SRO. The maximum amplitude of the zero temperature magnetization ($M_0$) occurs at $z = 3$ in the SRO, which is varied in the range $M_0 = [1.1, 1.8]\mu_B$. The magnetization profile at zero temperature is $M_z(1) = 0.3M_0$, $M_z(2) = 0.5M_0$, $M_z(3) = M_0$, $M_z(4) = 0.5M_0$.

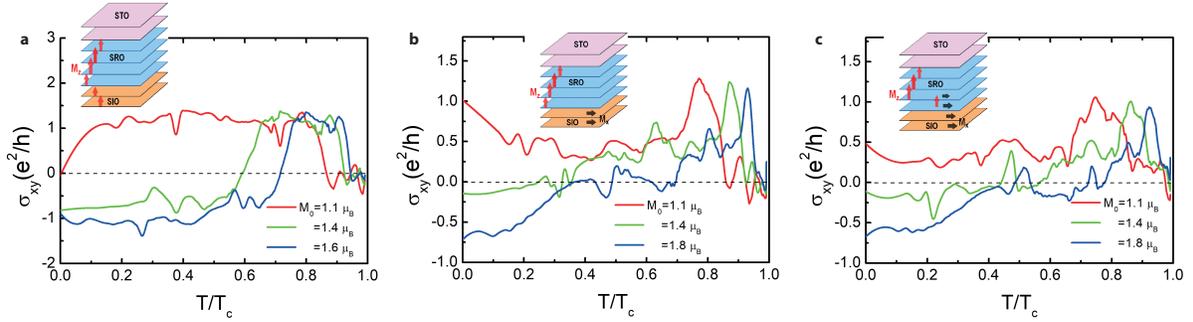

FIG. S16: **Induced magnetization on the Ir site.** $\sigma_{xy}$ versus $T/T_c$ for STO/SRO/SIO with an induced magnetization on the Ir site. **a**, The magnetization profile has only $z$-components and is not trivial in the SIO layers: $M_z(1) = 0.3M_0$, $M_z(2) = 0.5M_0$, $M_z(3) = M_0$, $M_z(4) = 0.5M_0$, $M_z(5) = M_z(6) = 0.1M_0$. **b**, The magnetization is parallel to the $c$-axis ($M_z$) in the SRO layers while it lies in the $ab$ plane ($M_x$) within the SIO layers close to the SRO/SIO interface. **c**, The magnetization is parallel to the $c$-axis in the SRO layers with a varying amplitude and an in-plane component develops in the first two SRO layers near the SRO/SIO interface and in the SIO side of the heterostructure. In (b) the magnetization profile at zero temperature is $M_z(1) = 0.3M_0$, $M_z(2) = 0.5M_0$, $M_z(3) = M_0$, $M_z(4) = 0.5M_0$, $M_z(5) = M_z(6) = 0$, and $M_x(5) = M_x(6) = 0.1M_0$. In (c) the layer magnetization profile at zero temperature is: $M_z(1) = 0.3M_0$, $M_z(2) = 0.5M_0$, $M_z(3) = M_0$, $M_z(4) = 0.5M_0$, $M_z(5) = M_z(6) = 0$, and $M_x(3) = 0.05M_0$, $M_x(4) = 0.1M_0$ and $M_x(5) = M_x(6) = 0.1M_0$.

magnetic anisotropy. In Fig. S16b, we show the AH conductivity with (b) the SIO magnetization parallel to $x$-axis, and (c) with a coexistence of out-of-plane and in-plane magnetic moments close to the SRO/SIO interface. The overall behavior is that the sign change is generally observed when $M_0$ is sufficiently large. For smaller Ru magnetic moments, the presence of an in-plane component at the interface tends to enhance the conductivity at low temperature (red curve in Fig. S16b). The addition of an out-of-plane magnetic component at the interface tends to shift the conductivity towards negative values.


\end{document}